\documentclass[journal]{IEEEtran}

\usepackage[linesnumbered,ruled]{algorithm2e}
\usepackage{graphicx,amssymb,mathrsfs,amsmath}
\usepackage{epstopdf}
\usepackage{subfigure}
\usepackage{epsfig}
\usepackage{setspace}
\usepackage{multirow}
\usepackage{booktabs}
\usepackage{soul}
\usepackage{color}
\usepackage{cite}
\usepackage{float}

\usepackage{lineno}

\usepackage{algpseudocode}
\usepackage{cite}

\makeatletter
\def\hlinew#1{%
  \noalign{\ifnum0=`}\fi\hrule \@height #1 \futurelet
   \reserved@a\@xhline}

\begin{document}
\setstcolor{red}
\title{Coupled Convolutional Neural Network with Adaptive Response Function Learning for Unsupervised Hyperspectral Super-Resolution}

\author{Ke Zheng,
        Lianru Gao,~\IEEEmembership{Senior Member,~IEEE,}
        Wenzhi Liao,~\IEEEmembership{Senior Member,~IEEE,}
        Danfeng Hong,~\IEEEmembership{Member,~IEEE,}
        Bing Zhang,~\IEEEmembership{Fellow,~IEEE,}
        Ximin Cui,
        and~Jocelyn Chanussot,~\IEEEmembership{Fellow,~IEEE}
        

\thanks{This work was supported by the National Natural Science Foundation of China under Grant No. 41722108 and No. 91638201, as well as with the support of the AXA Research Fund. (\emph{Corresponding author: Lianru Gao.}) }

\thanks{K. Zheng is with the Key Laboratory of Digital Earth Science, Aerospace Information Research Institute, Chinese Academy of Sciences, Beijing 100094, China, and also with the College of Geoscience and Surveying Engineering, China University of Mining and Technology (Bei Jing), Beijing 100083, China. (e-mail: zhengkevic@aircas.ac.cn)}
\thanks{L. Gao is with the Key Laboratory of Digital Earth Science, Aerospace Information Research Institute, Chinese Academy of Sciences, Beijing 100094, China. (e-mail: gaolr@aircas.ac.cn)}
\thanks{W. Liao is with the Sustainable Materials Management, Flemish Institute for Technological Research (VITO), 2400 Mol, Belgium, and also with the Image Processing and Interpretation, IMEC Research Group, Ghent University, 9000 Ghent, Belgium. (e-mail: wenzhi.liao@vito.be)}
\thanks{D. Hong is with the Remote Sensing Technology Institute (IMF), German Aerospace Center (DLR), 82234 Weßling, Germany. (e-mail: danfeng.hong@dlr.de)}
\thanks{B. Zhang is with the Key Laboratory of Digital Earth Science, Aerospace Information Research Institute, Chinese Academy of Sciences, Beijing 100094, China, and also with the College of Resources and Environment, University of Chinese Academy of Sciences, Beijing 100049, China. (e-mail: zb@radi.ac.cn)}
\thanks{X. Cui is with the College of Geoscience and Surveying Engineering, China University of Mining and Technology (Bei Jing), Beijing 100083, China. (e-mail: cxm@cumtb.edu.cn)}
\thanks{J. Chanussot is with the Univ. Grenoble Alpes, CNRS, Grenoble INP, GIPSA-lab, F-38000 Grenoble, France, also with Aerospace Information Research Institute, Chinese Academy of Sciences, Beijing 100094. (e-mail: 
jocelyn.chanussot@grenoble-inp.fr)}
}

\markboth{IEEE Transactions on Geoscience and Remote Sensing,~Vol.~XX, No.~XX, ~XXXX,~2020}%
{Shell \MakeLowercase{\textit{et al.}}: Coupled Convolutional Neural Network with Adaptive Response Function Learning for Unsupervised Hyperspectral Super-Resolution}
\maketitle
\begin{abstract}
Due to the limitations of hyperspectral imaging systems, hyperspectral imagery (HSI) often suffers from poor spatial resolution, thus hampering many applications of the imagery. Hyperspectral super-resolution refers to fusing HSI and MSI to generate an image with both high spatial and high spectral resolutions. Recently, several new methods have been proposed to solve this fusion problem, and most of these methods assume that the prior information of the Point Spread Function (PSF) and Spectral Response Function (SRF) are known. However, in practice, this information is often limited or unavailable. In this work, an unsupervised deep learning based fusion method – HyCoNet – that can solve the problems in HSI–MSI fusion without the prior PSF and SRF information is proposed. HyCoNet consists of three coupled autoencoder nets in which the HSI and MSI are unmixed into endmembers and abundances based on the linear unmixing model. Two special convolutional layers are designed to act as a bridge that coordinates with the three autoencoder nets, and the PSF and SRF parameters are learned adaptively in the two convolution layers during the training process. Furthermore, driven by the joint loss function, the proposed method is straightforward and easily implemented in an end-to-end training manner. The experiments performed in the study demonstrate that the proposed method performs well and produces robust results for different datasets and arbitrary PSFs and SRFs.
\end{abstract}
\graphicspath{{figures/}}
\begin{IEEEkeywords}
Hyperspectral Image, Super-Resolution, Coupled Convolutional Neural Network, Autoencoder, Adaptive Learning
\end{IEEEkeywords}
\section{Introduction}
\IEEEPARstart{A} hyperspectral image (HSI) is a data cube containing hundreds of contiguous narrow-bandwidth images covering a large wavelength range \cite{rasti2020feature}. Because of its high spectral resolution, HSI is very important in many applications such as land cover classification \cite{gao2014subspace,hong2020invariant,cao2020hyperspectral,cao2020an}, target detection \cite{guo2014weighted,li2018real,wu2019orsim,wu2019fourier}, feature extraction and dimensionality reduction \cite{he2016weighted,hong2017learning,xu2019superpixel,hong2019learning}, data fusion \cite{xu2019nonlocal,hu2019mima,hang2020classification}, and spectral unmixing \cite{tang2017integrating,hong2018sulora,yao2019nonconvex}. However, the hyperspectral imaging system often has a trade-off between spectral resolution and spatial resolution, due to hardware restrictions. This causes the spatial resolution of HSI to usually be coarser than that of multispectral imagery (MSI). The limited spatial resolution restricts further applications of HSI. To enhance the spatial resolution of HSI, a natural solution is to fuse it with higher-resolution MSI. This approach is called hyperspectral and multispectral image fusion (HSI-MSI Fusion). 

HSI-MSI Fusion is similar to the MSI pansharpening process in which a low spatial resolution MSI is fused with a high-resolution panchromatic (PAN) image. However, applying these pansharpening methods to directly fuse HSI and PAN imagery remains challenging as PAN images contain limited spectral information, and, as a result, spectral distortion can easily occur \cite{loncan2015hyperspectral}. Recently, HSI-MSI Fusion has attracted a lot of attention because the result preserves more accurate spectral information with a high-spatial resolution. The existing fusion methods can be categorized as: (1) extensions of pansharpening methods; (2) Bayesian-based approaches; and (3) matrix factorization-based methods \cite{hong2020learning}.

In the first category, Richard \emph{et al.} first attempted to use a pansharpening-based method to fuse HSI and MSI using a wavelet technique \cite{gomez2001wavelet}. However, the results were highly dependent on spectral resampling, which made it difficult to enhance the spatial resolution. Zhang \emph{et al.} proposed a pan-sharpening-based fusion method that used a 3-D wavelet transform \cite{zhang2007multi}. Chen \emph{et al.} proposed a framework for fusing HSI and MSI by dividing the HSI into several regions and fusing the HSI and MSI in each region using the pansharpening method \cite{chen2014fusion}. Aiazzi \emph{et al.} proposed a component substitution fusion method that took the spectral response function (SRF) as part of the model \cite{aiazzi2007improving}. Liu \emph{et al.} proposed a spectral preservation fusion method that was based on a simplified solar radiation and land-surface reflection model \cite{liu2000smoothing}.

In the second category, Eismann \emph{et al.} proposed a Bayesian-based fusion method that used a stochastic mixing model of the underlying spectral content to achieve resolution enhancement \cite{eismann2005hyperspectral}. Qi \emph{et al.} proposed a variational-based fusion method that assumed the target image in a low- dimensional subspace and that solved the fusion problem by alternating optimization with respect to the coding coefficients and the target image \cite{wei2015hyperspectral}. Simões \emph{et al.} formulated the fusion problem as a minimization of a convex objection containing two quadratic terms and an edge-preserving term \cite{simoes2014convex}. Akhtar \emph{et al.} proposed a non-parametric Bayesian sparse coding strategy which first inferred the probability distributions of the material spectra and then computed the sparse codes of the high-resolution image \cite{akhtar2015bayesian}.

Methods in the third category – the matrix factorization methods – usually assume that the HSI is composed of a series of pure spectral vectors and that the matrix HSI can be decomposed into abundances and endmembers. The fusion problem becomes an estimation problem for the endmembers of the low-resolution HSI and abundances of the high-resolution MSI. Kawakami \emph{et al.} proposed an unmixing approach to fusing a low-resolution HSI with a high-resolution RGB image. Firstly, the unmixing algorithm was employed to estimate the basis endmembers of the low-resolution HSI; this was then combined with a high-resolution RGB image to produce the final result \cite{kawakami2011high}. Instead of keeping the estimated endmembers of the low-resolution HSI fixed, Yokoya and Lanaras presented a coupled NMF (CNMF) to estimate endmembers and abundances using an alternating unmixing approach \cite{yokoya2011coupled, lanaras2015hyperspectral}. Wycoff \emph{et al.} restricted sparse regularization to abundances which assume that each pixel is composed of only a small number of endmembers \cite{wycoff2013non}. Akhtar \emph{et al.} proposed a sparse representation-based approach with a local spatial structure constraint whose main feature was the exploitation of the local patch prior knowledge using a greedy pursuit algorithm \cite{akhtar2014sparse}. Yi \emph{et al.} proposed an interactive feedback strategy fusion method with spectral unmixing and spatial constraints \cite{yi2018hyperspectral}. Tensor-based fusion methods utilizing tensor factorization with sparse constraint or subspace projection have also been proposed \cite{dian2017hyperspectral, xu2019nonlocal}.

In recent years, deep learning has been successfully applied in many spectral tasks \cite{long2017accurate, wu2018msri,jia2019collaborative, liang2018material}. To deal with the HSI–MSI fusion problem with deep learning, Wang \emph{et al.} proposed a convolutional neural network to fuse HSI and MSI using a residual network and a preserved spectral loss function \cite{wang2017deep}. Han \emph{et al.} proposed a partial densely connected network to fuse MSI and HSI spatial and spectral information \cite{han2018ssf}. Palsson \emph{et al.} proposed a 3-D convolutional neural network together with reducing the dimensionality of the HSI to make the fusion more computationally efficient \cite{palsson2017multispectral}. Dian \emph{et al.} initialized the HSI by solving a Sylvester equation and then implementing a neural network to learn the mapping between the initialized HSI and target HSI \cite{dian2018deep}. Xie \emph{et al.} proposed a fusion network that took the observation models of low-resolution images and the low-rank knowledge into consideration \cite{xie2019multispectral}. Wang \emph{et al.} proposed a deep learning-based blind hyperspectral image fusion method with iterative and alternating optimization strategy \cite{wang2019deep}. Han \emph{et al.} presented a multi-scale spatial and spectral fusion convolutional neural network (CNN) for HSI-MSI Fusion \cite{han2019multi}.
However, all of the deep learning-based methods mentioned above are supervised learning methods that are difficult to apply in practice because the high-resolution HSI needed for training is unavailable. Qu \emph{et al.} proposed an unsupervised deep learning-based fusion method with a sparse Dirichlet network \cite{qu2018unsupervised}. Zhou \emph{et al.} proposed a registration algorithm and a fusion algorithm to handle HSI and MSI image with significant scale difference and nonrigid distortion \cite{zhou2019integrated}. Fu \emph{et al.} proposed a camera spectral response (CSR) optimization layer to learn the spectral response with an unsupervised way \cite{fu2019hyperspectral}. 

Some of fusion methods assume that the prior information of the SRF or point spread function (PSF) is known. However, in practice, this information is often difficult to obtain \cite{simoes2014convex}. In this paper, an unsupervised deep learning-based fusion network that can handle situations where the PSF and SRF are unknown is proposed. The only information our proposed method requires is the spectral coverage of the MSI and HSI, which is easy to obtain from the data provider. To our knowledge, this is the first time which the unsupervised coupled CNN was developed with learnable PSFs for the HSI-MSI Fusion task. The main contributions of this study can be summarized as follows:
\begin{itemize}
\item A novel unsupervised network called HyCoNet is proposed to solve the HSI–MSI fusion problem for an unknown SRF and PSF. The results show that the proposed method can deal well with arbitrary SRFs and PSFs in comparison with nine state-of-the-art HSI–MSI fusion methods, as applied to four remote sensing datasets;
\item Based on the linear unmixing theory, three autoencoder networks are jointly coupled in the proposed method. Within these networks, the endmembers comprise the parameters of one convolution layer, which is shared between two autoencoder networks. Also, in order to improve the consistency of these networks, a learned PSF layer acts as a bridge connecting the low- and high-resolution abundances;
\item A joint-loss function that drives the model using unsupervised learning and an end-to-end training manner, thus providing a simple and direct training strategy for obtaining the final result, is introduced.
\end{itemize}

This paper is organized as follows. Section II describes the basic formulation of the HSI and MSI fusion problem. Section III introduces the proposed fusion model, including the network architecture and the joint-loss function. Section IV presents the experimental results and discussion; Section V is the conclusion.

\section{Problem Formulation}

The HSI-MSI fusion problem requires the estimation of the HSI, which has both high spectral and high spatial resolution and is denoted as $\mathbf{X} \in \mathbb{R}^{M \times N \times L}$, where $M$, $N$ and $L$ are the width, height, and number of spectral bands. The input images include a high spatial resolution MSI denoted as $\mathbf{Y} \in {\mathbb{R}^{M \times N \times l}}$ and a low spatial resolution HSI denoted as $\mathbf{Z} \in \mathbb{R}^{m \times n \times L}$, where $l$ is the number of spectral bands in $\mathbf{Y}$, and $m$ and $n$ are the width and height of $\mathbf{Z}$. In particular, $n \leqslant N$, $m \leqslant M$ and $l \leqslant L$. For convenience, we call the low spatial resolution HSI, high spatial resolution MSI and target HSI and low spatial resolution MSI as LrHSI, HrMSI and HrHSI, respectively. To simplify the notation in this section, we unfold the 3D data cube to form a 2D matrix: the 2D matrices of the three types of image are denoted as   $\mathbf{X} \in {\mathbb{R}^{MN \times L}}$, $\mathbf{Y} \in {\mathbb{R}^{MN \times l}}$ and $\mathbf{Z} \in {\mathbb{R}^{mn \times L}}$, respectively. 

\begin{figure}[!t]
    \centering
        \subfigure{
            \includegraphics[width=0.45\textwidth]{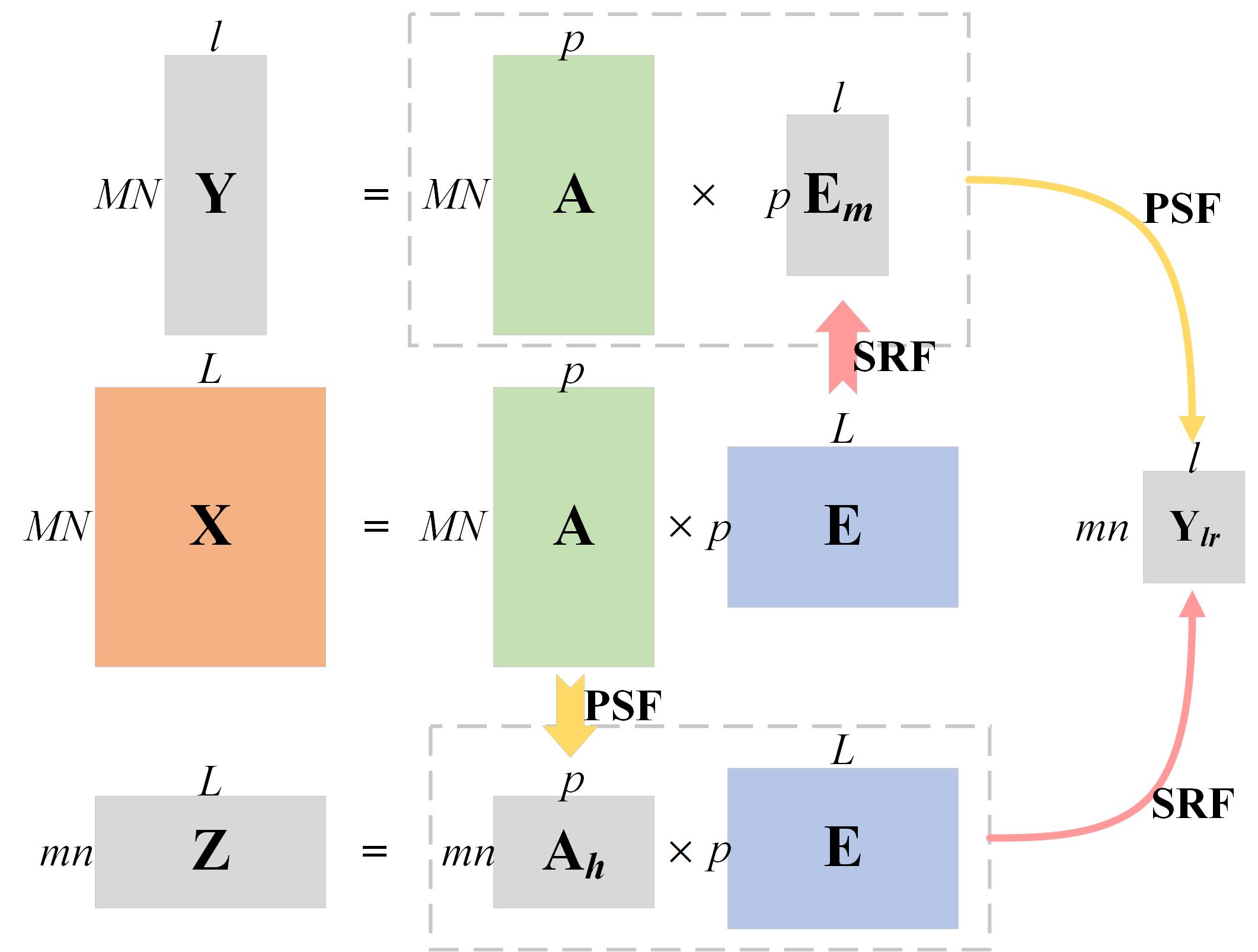}
        }
    \caption{The illustration of the relationship between the HrMSI, the LrHSI and the target HrHSI based on the linear unmixing model , where $\mathbf{Y}_{lr}$ represents low spatial resolution multispectral image (LrMSI).}
    \label{fig:fig1}
\end{figure}

The relationship between these images $\mathbf{X}$, $\mathbf{Y}$ and $\mathbf{Z}$ according to the hyperspectral linear unmixing model \cite{hong2018augmented} is shown in Fig. \ref{fig:fig1}. In the linear spectral unmixing model, each pixel of the HSI is assumed to be a linear combination of a set of pure spectral bases (called endmembers) and the coefficients of each pure spectral basis (called abundances). According to the linear spectral unmixing model, the target HrHSI can be described as: 
\begin{equation}
\label{equation1}
\begin{aligned}
	  \mathbf{X} = \mathbf{AE}
\end{aligned}
\end{equation}
where the matrix $\mathbf{A} \in\mathbb{R}^{MN \times p}$ is formed from the abundances, the matrix $\mathbf{E} \in \mathbb{R}^{p \times L}$ is made up of the endmembers, and $p$ is the number of pure spectral bases. This equation describes the degree of mixing for each pixel in the image $\mathbf{X}$.

Similarly, the input $\mathbf{Z}$ can also be expressed as a linear combination of the same endmembers $\mathbf{E}$:
\begin{equation}
\label{equation2}
\begin{aligned}
	  {\mathbf{Z}} = {{\mathbf{A}}_h}{\mathbf{E}} = {\mathbf{S}} * {\mathbf{X}} = {\mathbf{S}} * {\mathbf{AE}}
\end{aligned}
\end{equation}
where the matrix ${{\mathbf{A}}_h} \in {\mathbb{R}^{mn \times p}}$ represents the abundances, the matrix ${\mathbf{S}} \in {\mathbb{R}^{{L_m} \times {L_n}}}$ is the point spread function (PSF), which describes the spatial degradation function, and $*$ denotes the convolution operator \cite{kang2020learning}. $L_m$ and $L_n$ are the spatial size of the convolution filter. $\mathbf{Z}$ also denotes the spatially degraded version of image $\mathbf{X}$. The input $\mathbf{Y}$ is the spectrally degraded version of $\mathbf{X}$:
\begin{equation}
\label{equation3}
\begin{aligned}
	  \mathbf{Y} = \mathbf{X}{\mathbf{R}} = {\mathbf{AER}}
\end{aligned}
\end{equation}
where the matrix ${\mathbf{R}} \in {\mathbb{R}^{L \times l}}$ is the spectral response function (SRF), which describes the spectral degradation process.

Moreover, the spectral degraded version of $\mathbf{Z}$ should approximate to the spatially degraded version of $\mathbf{Y}$:
\begin{equation}
\label{equation4}
\begin{aligned}
	  \mathbf{Y}_{lr} = {\mathbf{Z}}{\mathbf{R}} = {\mathbf{S}} * {\mathbf{Y}}
\end{aligned}
\end{equation}
where the matrix $\mathbf{Y}_{lr}$ $\in {\mathbb{R}^{m \times n \times l}}$ represents the low spatial resolution multispectral image (LrMSI).

In addition, our goal is to estimate $\mathbf{X}$ using the inputs $\mathbf{Y}$ and $\mathbf{Z}$ with the following constraints also satisfied:
\begin{equation}
\label{equation5}
\begin{aligned}
	  \begin{gathered}
  \sum\limits_{j = 1}^p {{a_{ij}}}  = 1\;\;\forall \;i,j \hfill \\
  {a_{ij}} \geqslant 0\;\;\forall \;i,j \hfill \\
  1 \geqslant {e_{ij}} \geqslant 0\;\;\forall \;i,j \hfill \\ 
\end{gathered} 
\end{aligned}
\end{equation}
where $a_{ij}$ is a component unit of $\mathbf{A}$ and $e_{ij}$ is a component unit of $\mathbf{E}$. These constraints relate to the sum-to-one property of the abundance, the non-negative property of the abundance and the bounded non-negative property of the endmembers, respectively \cite{wei2015hyperspectral}. In addition, the abundances should be sparse, meaning that each HSI pixel is composed of only a few pure spectral bases.

\section{Proposed Method}

According to Eqs. (\ref{equation1})-(\ref{equation3}), to solve the HSI–MSI fusion problem, the key point is to estimate the high spatial resolution abundance matrix $\mathbf{A}$ and the spectral bases matrix $\mathbf{E}$. The HrMSI provides detailed spatial contextual information that is highly correlated with $\mathbf{A}$. Also, the LrHSI preserves the spectral information, which is highly consistent with the target spectral endmembers matrix $\mathbf{E}$. The basic idea of the proposed method is based on matrix factorization. The proposed HyCoNet is an unsupervised network that includes three coupled autoencoder networks. The target HrHSI is embedded in one of the networks – this will be elaborated on in the part A of this section, and the Part B introduces the joint objective function used in the training.

\subsection{Coupled Autoencoder Network for Image Fusion}\label{couplednet}

\begin{figure*}[!t]
	  \centering
		\subfigure{
			\includegraphics[width=0.9\textwidth]{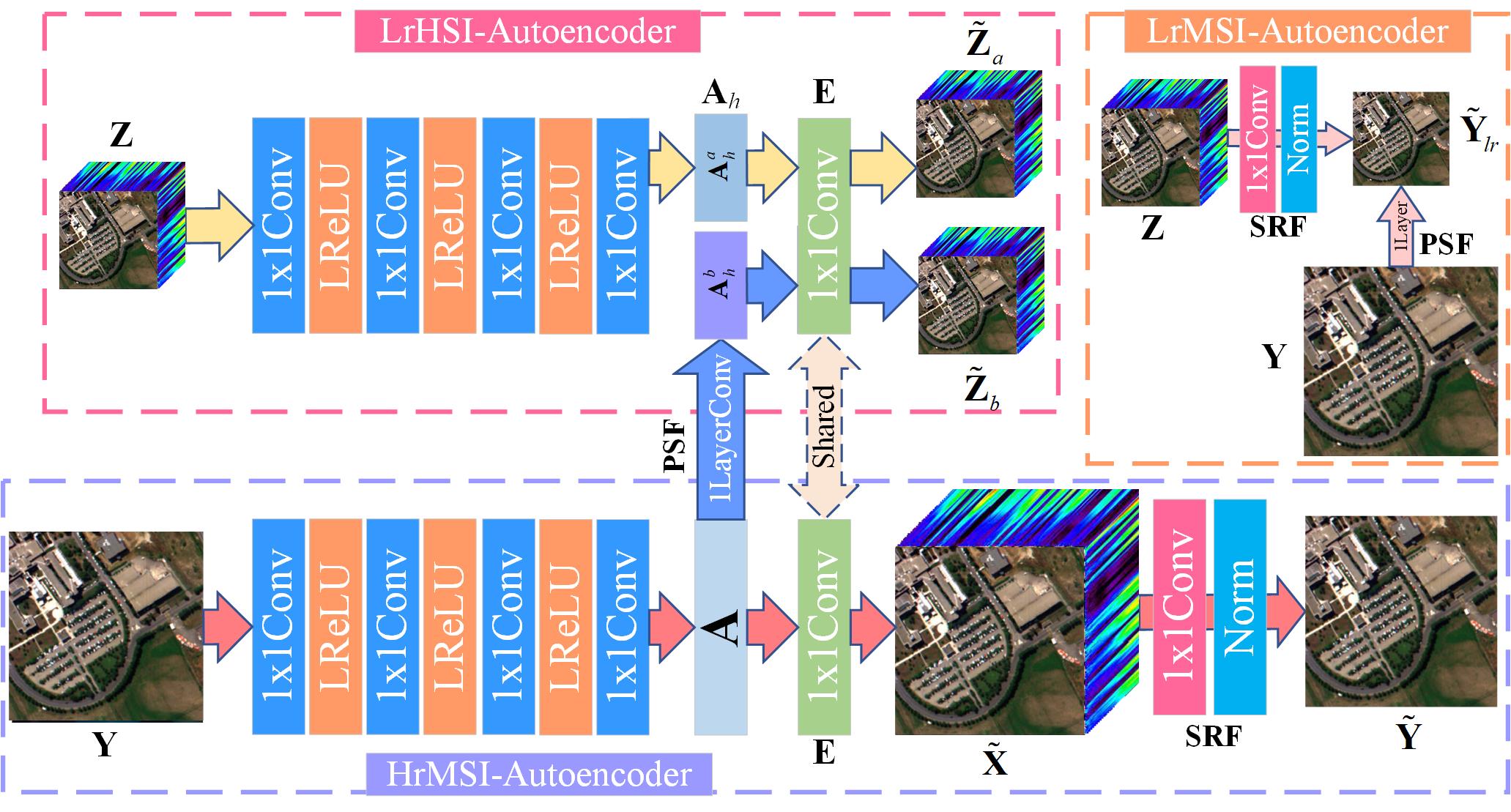}
		}
        \caption{The generator networks of proposed HyCoNet method which contains three coupled autoencoder nets: LrHSI-Autoencoder, HrMSI-Autoencoder and LrMSI autoencoder.}
\label{fig:fig2}
\end{figure*}

The proposed network is composed of three autoencoder nets, as shown in Fig. \ref{fig:fig2}. The upper one called LrHSI autoencoder, and the lower one called HrMSI autoencoder. The estimated target HrHSI $\mathbf{X}$ is embedded in the HrMSI autoencoder. The LrMSI autoencoder can be seen at the upper right.

Since the 2D convolution layer, which has a kernel size of $1\times1$, is equivalent to the fully connected layer when it applied to a spectral vector, all the fully connected layers in the traditional autoencoder network are replaced by the convolution layers, as shown in Fig. \ref{fig:fig2}. In addition, all the convolution kernel sizes are set to be $1\times1$, except for the PSF convolution layer. Instead of using the fully connected layer, the convolution layer is used to preserve the spatial structure of the input image cube for easy implementation of the PSF operation. Further details of the PSF operation are discussed below.

In the LrHSI autoencoder, the network tries to learn an approximation to the identity function $f({\mathbf{Z}}) \approx \mathbf{Z}$. Since the input LrHSI contains sufficient spectral information, the endmembers $\mathbf{E}$ and abundances $\mathbf{A}_h$ can be extracted during the reconstruction process for this autoencoder.

The module before the latent variable ${\mathbf{A}}_h^a$ is the encoder:
\begin{equation}
\label{equation6}
\begin{aligned}
	  {\mathbf{A}}_h^a = {f_{en}}(\mathbf{Z})
\end{aligned}
\end{equation}
where the $f_{en}()$ try to learn a nonlinear mapping which transforms the input LrHSI to its abundances $\mathbf{A}_h^a$; ${\mathbf{A}}_h^a \in {\mathbb{R}^{m \times n \times p}}$ represents the abundances of $\mathbf{Z}$, and $p$ is the number of endmembers.

The module after the latent variable is the decoder function:
\begin{equation}
\label{equation7}
\begin{aligned}
	  {{\mathbf{\tilde Z}}_a} = {f_{de}}({\mathbf{A}}_h^a)
\end{aligned}
\end{equation}
where ${{\mathbf{\tilde Z}}_a}$ is the output of this upper autoencoder, which represents the reconstructed input image cube. The decoder function ${f_{de}}()$ is a convolution layer without bias and is shown as the green convolution layer “1x1 Conv” in Fig. \ref{fig:fig2}. This layer is also used in the HrMSI autoencoder, meaning that these two autoencoders share the same parameters as this convolution layer. The parameters of the shared convolution layer are denoted as the endmembers ${\mathbf{E}} \in {\mathbb{R}^{p \times 1 \times 1 \times L}}$, where $1\times1$ represents the spatial size of this convolution kernel. The size of each convolution kernel is $p\times1\times1$ and the number of kernels is $L$.

The reason, that we call the latent variables ${\mathbf{A}}_h^a$ and parameters $\mathbf{E}$ as the abundances and endmembers, is that they are trying to reconstruct $\mathbf{Z}$ by matrix multiplication: ${{\mathbf{\tilde Z}}_a} = \mathbf{A}_h^a\mathbf{E}$. This is the same idea as matrix decomposition except that the process is optimized by gradient descent in neural network. In order to satisfy the non-negativity restrictions for $\mathbf{A}_h$ and $\mathbf{E}$, several tricks are applied, as described in more detail in Part \ref{jointlossfunc}.

The structure of the HrMSI autoencoder is similar to that of the LrHSI autoencoder and includes an encoder function $h_{en}()$ and a decoder function $h_{de}()$. The encoder function can be expressed as:
\begin{equation}
\label{equation8}
\begin{aligned}
	  {\mathbf{A}} = {h_{en}}(\mathbf{Y})
\end{aligned}
\end{equation}
where $\mathbf{A}$ is the high-resolution abundance, $h_{en}()$ is the encoder function and $\mathbf{Y}$ is the input {HrMSI}.

The decoder for the HrMSI autoencoder consists of two parts – a shared convolution layer and the SRF:
\begin{equation}
\label{equation9}
\begin{aligned}
	  \mathbf{\tilde Y} = {h_{de}}({\mathbf{A}}) = SRF({f_{de}}({\mathbf{A}}))
\end{aligned}
\end{equation}
where ${{\mathbf{\tilde Y}}}$ is the reconstructed HrMSI $\mathbf{Y}$, $h_{de}()$ is the decoder function, $f_{de}()$ is the shared convolution layer containing the parameters of the endmember matrix $\mathbf{E}$, and $SRF()$ is the spectral resampling operation. The HrHSI is the output of the shared convolution layer:
\begin{equation}
\label{equation10}
\begin{aligned}
	  \mathbf{\tilde X} = {f_{de}}({\mathbf{A}})
\end{aligned}
\end{equation}
where ${\mathbf{\tilde X}}$ is the estimated target image.

To handle the situation where the SRF parameters are unknown, a convolution layer and a normalization layer are placed after the target ${\mathbf{\tilde X}}$ to learn the unknown parameters of the SRF. The SRF consists of spectral resampling from HSI to MSI and the process can be defined as:
\begin{equation}
\label{equation11}
\begin{aligned}
	  {\varphi _i} = \frac{{\int_{{\lambda _{i,L}}}^{{\lambda _{i,U}}} {R(\lambda )\varepsilon (\lambda )d\lambda } }}{{\int_{{\lambda _{i,L}}}^{{\lambda _{i,U}}} {R(\lambda )d\lambda } }}
\end{aligned}
\end{equation}
where ${\varphi_i}$ is the spectral radiance of band $i$ of the HrMSI, $\lambda $ is the wavelength, ${\lambda _{i,U}}$ and ${\lambda _{i,L}}$ are the wavelength bounds of band $i$ of the HrMSI, $R()$ is the spectral response function, and $\varepsilon $ is the spectral radiance of the HrHSI. To implement the spectral resampling in the neural network, a convolution layer with kernel size $1 \times 1$ (shown in red and labeled “1x1Conv” in Fig. \ref{fig:fig2}) is added after the target image ${\mathbf{\tilde X}}$ to simulate the numerator of Eq. \ref{equation11}. A normalization layer (labeled “Norm” in Fig. \ref{fig:fig2}) follows the convolution layer and simulates the denominator of Eq. \ref{equation11}. 

Therefore, the SRF process within our network can be expressed as:
\begin{equation}
\label{equation12}
\begin{aligned}
	  {\varphi _i} = SRF({\varepsilon _\lambda }) = \frac{{\sum\limits_{\lambda  = {\lambda _{i,L}}}^{{\lambda _{i,U}}} {{w_{i,\lambda }}{\varepsilon _\lambda }} }}{{\sum\limits_{\lambda  = {\lambda _{i,L}}}^{{\lambda _{i,U}}} {{w_{i,\lambda }}} }}
\end{aligned}
\end{equation}
where ${\varphi _i}$ is the band $i$ image in ${{\mathbf{\tilde Y}}}$, ${w_{i,\lambda }}$ is the weight of the SRF convolution layer, and ${\varepsilon _\lambda }$ is the band with wavelength $\lambda $ in ${\mathbf{\tilde X}}$. The use of this function means that the convolution layer and the normalization layer integrate the HrHSI $\mathbf{\tilde X}$ with the weights of the convolution layer between the upper and lower spectral bounds.  It also means that our network assumes the spectral coverage of HrMSI bands is known and that each convolution kernel of the convolution layer only covers the spectral range corresponding to the bands in $\mathbf{\tilde X}$. The number of convolution kernels is equals to the number of bands in $\mathbf{\tilde Y}$.

The PSF means that a given pixel is a weighted combination of contributions from the pixel and its neighboring pixels \cite{wang2017effect, hong2015novel}. In the fusion problem, this means that a pixel in the LrHSI is a weighted combination of local pixels from the HrHSI. Therefore, the PSF is a convolution process and a convolution operation can easily be implemented as part of the convolutional neural network. In our network, to simulate the PSF, a convolution layer with 1 input channel and 1 output channel is implemented for every band of the abundance ${\mathbf{A}}$. According to the definition of the PSF, the relationship between $\mathbf{X}$ and $\mathbf{Z}$ can be expressed as $\mathbf{Z} = PSF(\mathbf{X})$. Also, ${\mathbf{X}}$ and ${\mathbf{Z}}$ are composed of the same linear unmixing endmembers. Therefore, another low-resolution abundance ${\mathbf{A}}_h^b$ can be characterized as:
\begin{equation}
\label{equation13}
\begin{aligned}
	  {\mathbf{A}}_h^b = PSF({\mathbf{A}})
\end{aligned}
\end{equation}
where $PSF()$ is the convolution layer labeled with the blue arrow and “1LayerConv” in Fig. \ref{fig:fig2}. The spatial size of this PSF convolution kernel is the same as the ratio of the Ground Sampling Distance (GSD) of the LrHSI to that of the HrMSI, and the stride of the convolution layer equals the kernel size. The PSF process acts as a bridge between the LrHSI autoencoder and the HrMSI autoencoder and forces the reconstructed image to be spectrally consistent ${\mathbf{\tilde X}}$. Therefore, another LrHSI ${\mathbf{\tilde{Z}}}_b$ can be reconstructed using $\mathbf{A}_h^b$ and $\mathbf{E}$:
\begin{equation}
\label{equation13plus}
\begin{aligned}
	  {\mathbf{\tilde{Z}}}_b = f_{de}(\mathbf{A}_h^b) 
\end{aligned}
\end{equation}

In addition, the spatial degraded version of the HrMSI is equivalent to the spectrally degraded version of the LrHSI shown as “LrMSI- Autoencoder” in Fig. \ref{fig:fig2}. This relation can be expressed as:
\begin{equation}
\label{equation14}
\begin{aligned}
	  PSF(\mathbf{Y}) = {\mathbf{\tilde Y}}_{lr}^a \approx {\mathbf{\tilde Y}}_{lr}^b = SRF(\mathbf{Z})
\end{aligned}
\end{equation}
where ${\mathbf{\tilde Y}}_{lr}$ is the estimated LrMSI.

\subsection{Joint Loss Function}\label{jointlossfunc}

\begin{figure}[!t]
    \centering
        \subfigure{
            \includegraphics[width=0.45\textwidth]{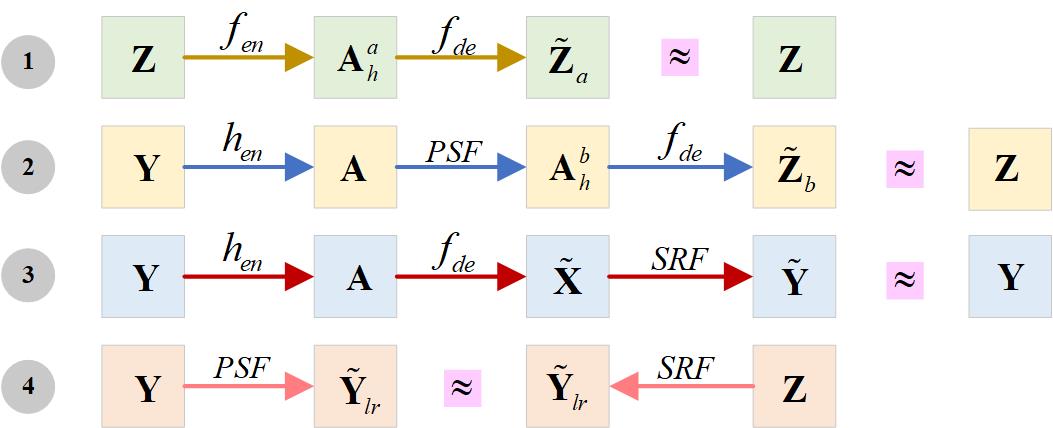}
        }
    \caption{The illustration of flow charts for the inputs $\mathbf{Y}$, $\mathbf{Z}$ and estimated target $\mathbf{\tilde{X}}$.}
    \label{fig:fig3}
\end{figure}

Fig. \ref{fig:fig3} shows flow charts for the proposed method, including input images, output images and corresponding target images. The first row of Fig. \ref{fig:fig3} indicates the inference of the input $\mathbf{Z}$ in the LrHSI-Autoencoder and is also represented by the yellow arrows in Fig. \ref{fig:fig2}. The inference of HrMSI $\mathbf{Y}$ includes two parts. The first one is shown as the second row of Fig. \ref{fig:fig3}, which includes the PSF operation to generate the LrHSI – this process is represented by the blue arrows Fig. \ref{fig:fig2}. The second part can be seen in the third row of Fig. \ref{fig:fig3} and consists of the SRF operation; this process is represented by the red arrows in Fig. \ref{fig:fig2}. The last row of Fig. \ref{fig:fig3} shows the relationship between the two LrMSIs. Therefore, the objective function for reconstruction can be expressed as:
\begin{equation}
\label{equation15}
\begin{aligned}
\begin{gathered}
\begin{array}{l}
{L_{base}}({\bf{Y}},{\bf{Z}}) = {\left\| {{\bf{Z}} - {\bf{\tilde Z}}_a^{}} \right\|_1} + \alpha {\left\| {{\bf{Z}} - {\bf{\tilde Z}}_b^{}} \right\|_1}\\
\;\;\;\;\;\;\;\;\;\;\;\;\;\;\;\;\; + \beta {\left\| {{\bf{Y}} - {\bf{\tilde Y}}} \right\|_1} + \gamma {\left\| {{\bf{\tilde Y}}_{lr}^a - {\bf{\tilde Y}}_{lr}^b} \right\|_1}
\end{array}
\end{gathered} 
\end{aligned}
\end{equation}
where $\alpha$, $\beta$ and $\gamma$ are trade-off parameters that tune the weights between these reconstruction errors.

The sum-to-one and non-negative properties given in Eq. \ref{equation5} also need to be satisfied. First, we constrain the sum of the abundances in the channel dimension to meet the sum-to-one property:
\begin{equation}
\label{equation16}
\begin{aligned}
	 \begin{gathered}
  {L_{sum2one}}(\mathbf{Y},\mathbf{Z}) = {\left\| {1 - \sum\limits_{i = 1}^p {{{\mathbf{A}}_i}} } \right\|_1} \hfill \\
  \;\;\;\;\;\;\;\;\;\;\;\;\;\;\;\;\;\;\;\;\;\;\; + {\left\| {1 - \sum\limits_{i = 1}^p {{\mathbf{A}}_{h,i}^a} } \right\|_1} + {\left\| {1 - \sum\limits_{i = 1}^p {{\mathbf{A}}_{h,i}^b} } \right\|_1} \hfill \\ 
\end{gathered}
\end{aligned}
\end{equation}
where $i$ indicates the $i$th band of the abundance matrix $\mathbf{A}$. Although the softmax function can be used to strictly enforce the sum-to-one property for the abundances, the resulting convergence accuracy is lower than for the proposed method. Part \ref{constraintfuc} will explain this in detail.

\begin{table*}[!t]
\centering
\caption{The used data in the experiment.}
\begin{tabular}{ccccc}
\hline
\hline
      & Pavia University & Indian Pines & Washington DC & University of Houston \\
\hline
Spatial size of HrHSI & 336x336 & 144x144 & 304x304 & 320x320 \\
Spectral range of HrHSI & 466-834nm & 400-2500nm & 400-2500nm & 403-1047nm \\
Number bands of HrHSI & 103   & 191   & 191   & 46 \\
GSD ratio & 4     & 4     & 8     & 8 \\
Spatial size of LrHSI & 84x84 & 36x36 & 38x38 & 40x40 \\
Bands of HrMSI & Blue-Green-Red & Blue to SWIR2 & Blue to SWIR2 & Blue-Green-Red \\
Number bands of HrMSI & 3     & 6     & 6     & 3 \\
\hline
\hline
\end{tabular}%
\label{tab:table1}
\end{table*}

Secondly, to enforce the non-negative property, several tricks are applied during training. The clamp function is applied to the output of the last convolution layer of both the encoder nets and decoder nets to force all the elements of the abundances and reconstruction images into the range [0,1]. Although we tried to use the sigmoid activation layer for this purpose, we found that it was difficult to make the network converge using this function. In addition, the weights of the shared convolution layer (containing the endmember parameter matrix, $\mathbf{E}$), PSF layer and SRF layer should also meet the non-negative property. Since the weights of these layers may be updated to a negative value after the back-propagation, we applied the clamp function to these layers after the weights were updated to force these weights into the non-negative range. The range of the clamp function was [0, 1]. As a result, the weights satisfied the non-negative and bounding constraints for each forward propagation, except for the first time.

Since each pixel of the HSI is composed of a small number of pure spectral bases, the abundance matrix $\mathbf{A}$ should be sparse. To guarantee the sparsity of the abundance, the KL divergence is used to ensure that most of the elements in the abundance are close to a small number:
\begin{equation}
\label{equation17}
\begin{aligned}
	 \begin{gathered}
  {L_{sparse}}(\mathbf{Y},\mathbf{Z}) = \sum\limits_{i = 1}^s {\sum\limits_{j = 1}^p {KL(a\parallel {{\tilde a}_{i,j}})} }  \hfill \\
  \;\;\;\;\;\; = \sum\limits_{i = 1}^s {\sum\limits_{j = 1}^p {(a\log (\frac{a}{{{{\tilde a}_{i,j}}}}) + (1 - a)\log (\frac{{1 - a}}{{1 - {{\tilde a}_{i,j}}}}))} }  \hfill \\ 
\end{gathered}
\end{aligned}
\end{equation}
where $s$ is the number of pixels, $p$ is the number of convolution kernels and also the number of endmembers, $a$ is a sparsity parameter which is set to a small value close to zero (0.0001 in our network), and ${\tilde a_{i,j}}$ is an element of the abundance matrix $\mathbf{A}$. To satisfy this constraint, the elements of $\mathbf{A}$ must mostly be near zero. The sparsity constraint is also applied to the abundance matrix $\mathbf{A}_h^a$.

Ultimately, our aim is to solve the fusion problem in accordance with the optimization problem:
\begin{equation}
\label{equation18}
\begin{aligned}
	 \begin{gathered}
  L(\mathbf{Y},\mathbf{Z}) = {L_{base}}(\mathbf{Y},\mathbf{Z}) \hfill \\
  \;\;\; + \mu {L_{sum2one}}(\mathbf{Y},\mathbf{Z}) + \nu {L_{sparse}}(\mathbf{Y},\mathbf{Z}) \hfill \\ 
\end{gathered}
\end{aligned}
\end{equation}
where $\mu$ and $\nu$ are the trade-off parameters used to balance the errors. This loss function can be directly used in the optimizer, thus providing a simple and direct solution.

\section{Experiments}

\begin{figure*}[!ht]
        \subfigure[Parameters disscussion for $\alpha$ and $\beta$.]{
		    \centering
			\includegraphics[height=5.2cm]{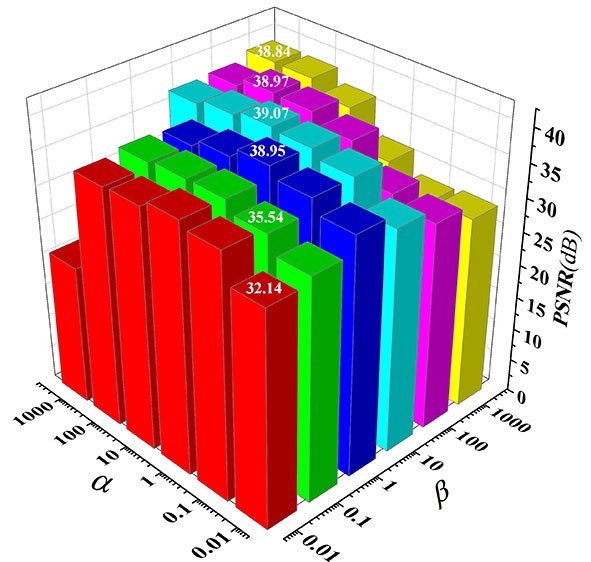}
            \label{fig:fig4}
        }\hfill
		\subfigure[PSNR results using different $\gamma$]{
		    \centering
			\includegraphics[height=4.6cm]{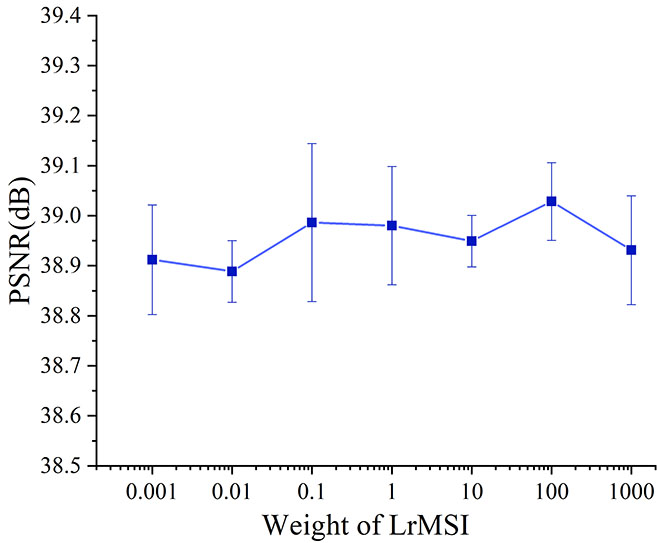}
            \label{fig:fig5}
		}\hfill
		\subfigure[Accuracy comparison between $\mu$ and $\nu$]{
		    \centering
		    \includegraphics[height=5.2cm]{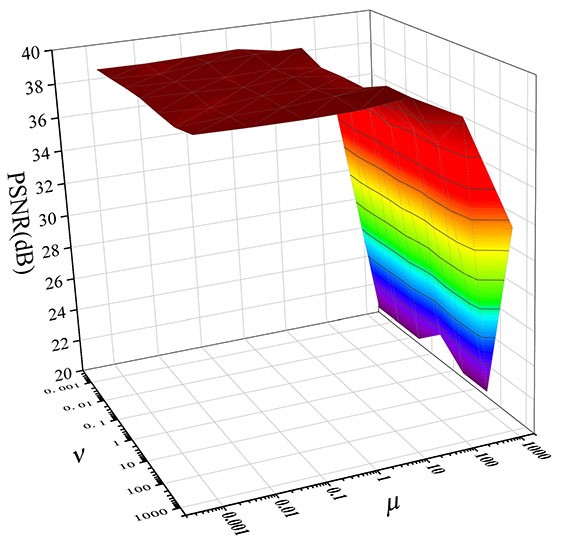}
		    \label{fig:fig6}
		}
\label{fig:parameters-disscussion}
\caption{Parameters disscussion. (a): Fusion accuracy of using different $\alpha$ and $\beta$  in Pavia University data, where $\alpha$ and $\beta$ mean the weights of reconstruction errors for LrHSI and HrMSI, respectively; (b): PSNR results of the different $\gamma$ in Pavia University data, where $\gamma$ represents the weight of LrMSI reconstruction error; (c): Accuracy comparison between $\mu$ and $\nu$ in the Pavia University data. $\mu$ and $\nu$ are the weights of sum-to-one error and sparsity error.}
\end{figure*}

\begin{figure}[!ht]
    \centering
    \includegraphics[height=7cm]{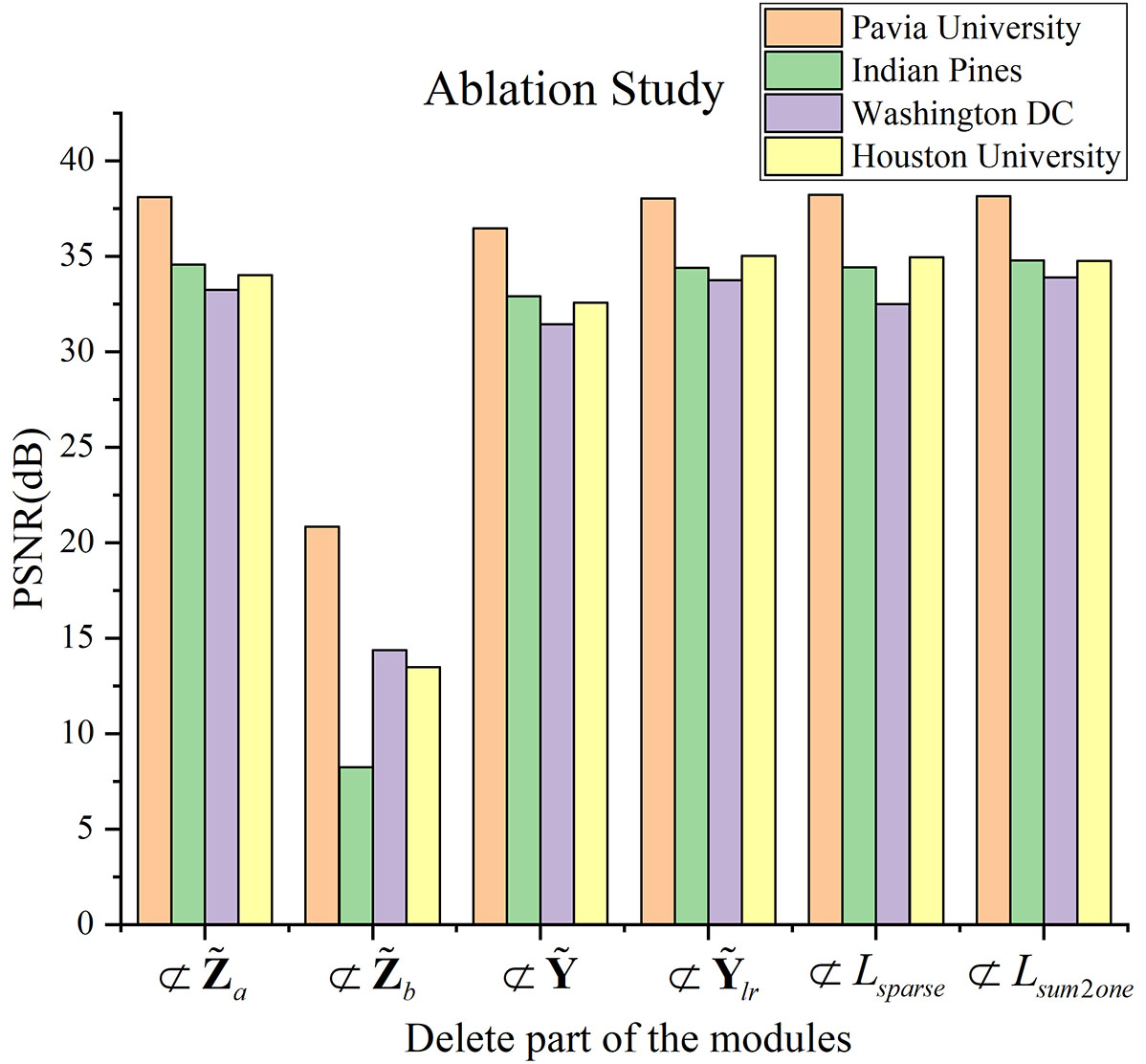}
    \caption{Ablation study for the proposed network. $\not\subset$ means removing part of the network/losses and studying it's performances.}
    \label{fig:fig6-2}
\end{figure}

\begin{figure}[!ht]
        \subfigure[Convergence curve using clamp and softmax function]{
		    \centering
			\includegraphics[height=3.5cm]{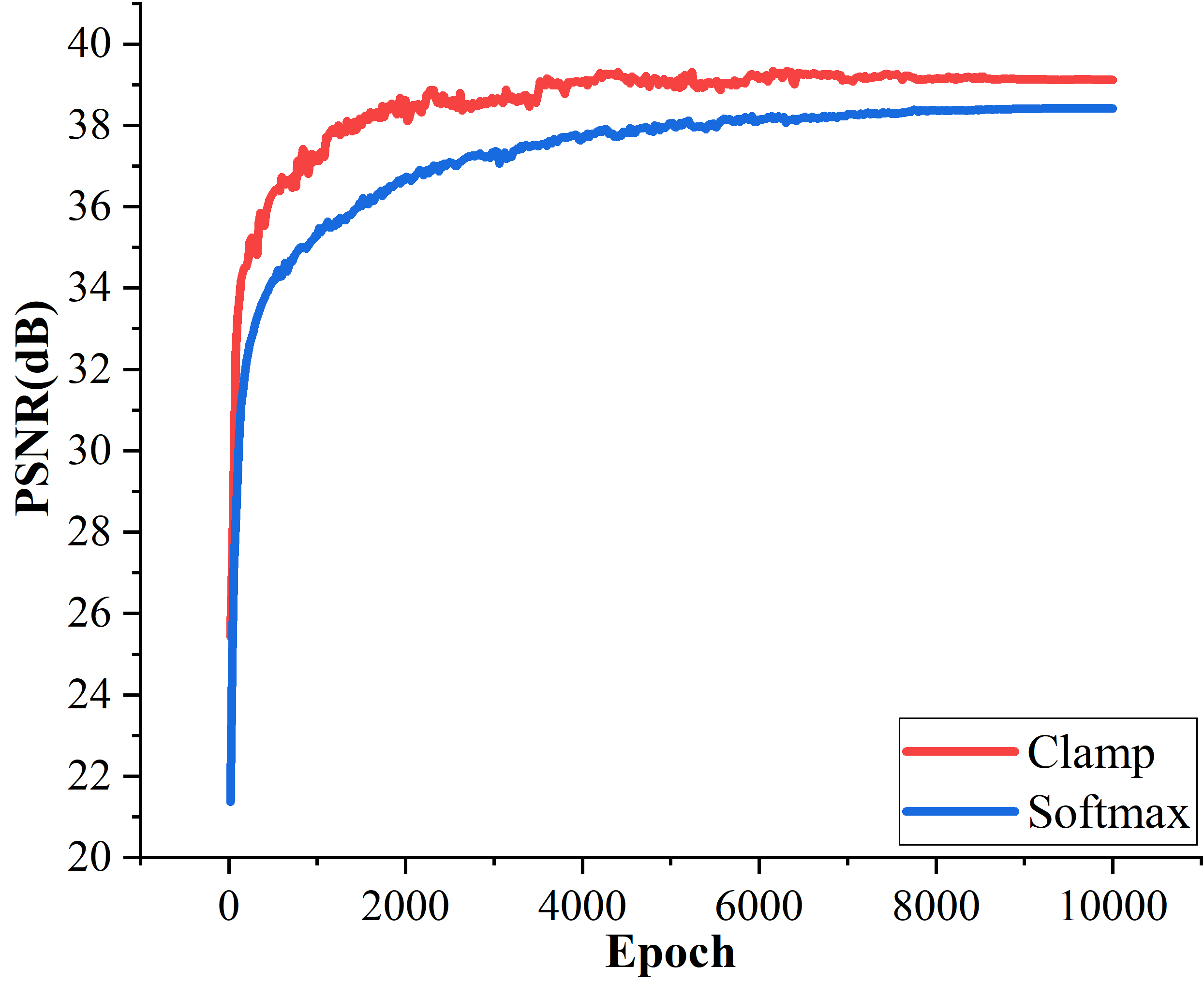}
            \label{fig:fig7a}
        }\hfill
		\subfigure[Function curves of clamp and softmax function]{
		    \centering
			\includegraphics[height=3.5cm]{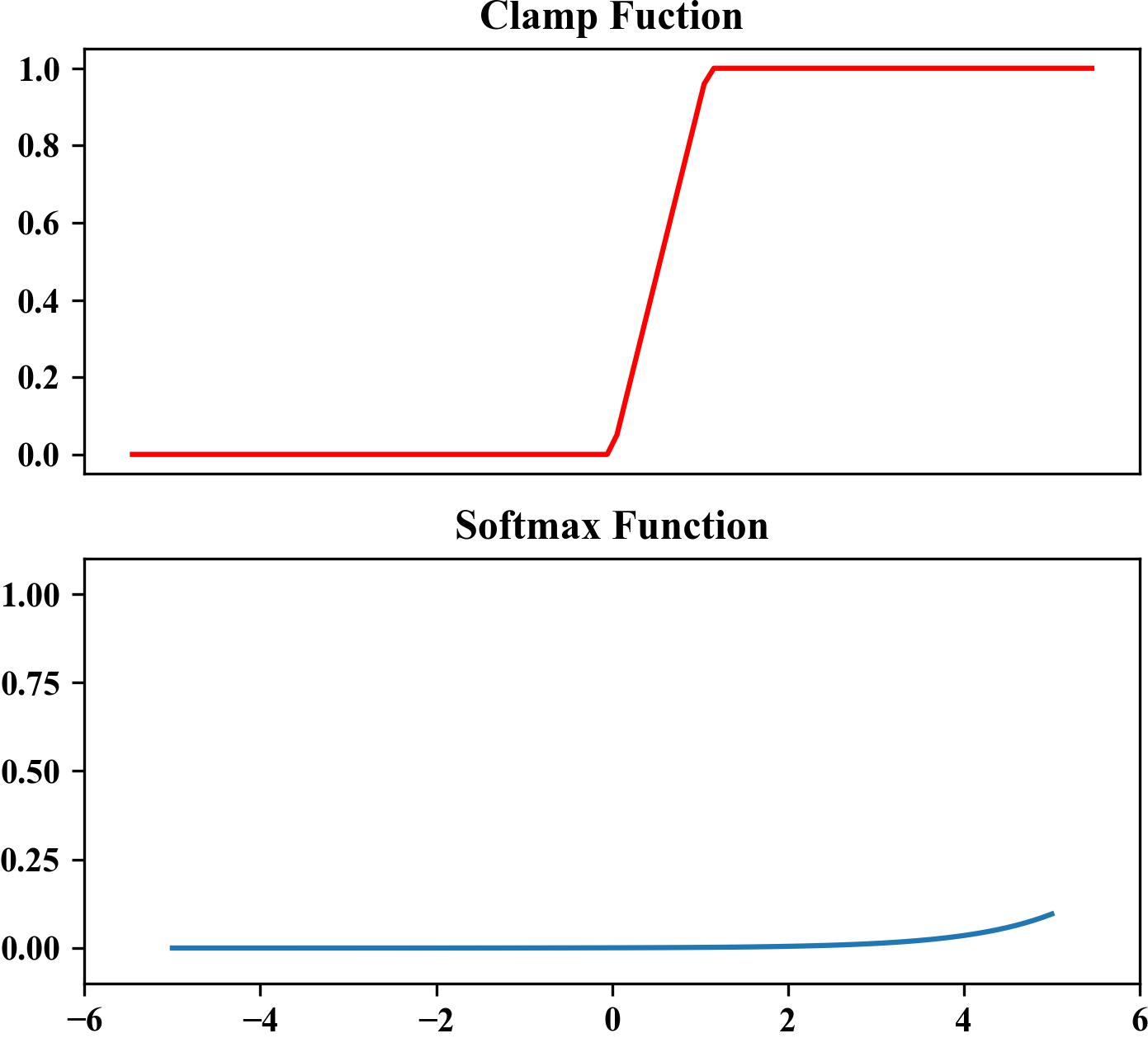}
            \label{fig:fig7b}
		}\hfill
\label{fig:ConstraintFunction}
\caption{Convergence curve when using different constrained functions for abundances. (a) Convergence accuracy for comparing clamp function and softmax function, respectively. (b) Function curves of clamp and softmax function for a vector evenly spaced 0.01 over the range [-5, 5].}
\end{figure}

To get an accurate assessment of the fusion quality and evaluate the performances of different fusion methods, simulation experiments \cite{yokoya2017hyperspectral} are used in the experiment.
 The proposed HyCoNet was implemented using four different simulated data sets. Firstly, the sensitivity of the trade-off parameters $\alpha$ , $\beta$ , $\gamma$ , $\mu$ and $\nu$ was evaluated. Secondly, the constraints on the abundances were investigated. Thirdly, we compared the effectiveness of the method for different numbers of endmembers. Fourthly, the learned PSF kernels were investigated using different spatially down-sampled kernels. The character of the estimated abundances was then explored. Finally, the fused images obtained using the different methods were evaluated using both visual and quantitative measures.

\subsection{Experimental Dataset}

The proposed HyCoNet was evaluated using four widely used HSI datasets: Pavia University, Indian Pines, Washington DC, and University of Houston. The Pavia University data were acquired by the ROSIS-3 optical airborne sensor in 2003. This image consists of $610\times340$ pixels with a GSD of 1.3m and spectral range of 430 nm–840 nm in 115 bands. Due to the effects of noise and water vapor absorption, 12 bands have been removed. An area covering $366\times366$ pixels in the lower-left corner of the image and containing 103 bands was selected for use in this experiment. The Indian Pines data were acquired by the AVIRIS sensor in 1992. This image consists of $145\times145$ pixels with a 20m GSD; the spectral range is 400 nm–2500 nm covering 224 bands. After removing 33 noisy bands, we selected a $144\times144$-pixel area with 191 bands as experimental data. The Washington DC data were acquired by the HYDICE sensor in 1995. This image has an area of $1280\times307$ pixels and a GSD of 2.5m. The spectral range is 400 nm–2500 nm consisting of 210 bands. After removing 19 noisy bands, we selected 191 bands covering $304\times304$ pixels for use. The University of Houston data were used in the 2018 IEEE GRSS Data Fusion Contest \cite{le20182018}, and consist of $601\times2384$ pixels with a 1-m GSD. The data covers the spectral range 380 nm–1050nm with 48 bands. We selected 46 bands consisting of $320\times320$ pixels from this imagery for use as experimental data.

\subsection{Implementation Details}

In the experiment, simulated spatial downsampling was used to generate the LrHSI using a Gaussian filter, as is widely used in remote sensing \cite{hong2019learnable}. In the experiment, the width and height of the Gaussian filter was set equal to the ratio between the high-resolution GSD and the low-resolution GSD. The standard deviation of all the Gaussian filters was set to 0.5, except as described in part \ref{learnedpsf} and \ref{comparesoa} where different standard deviations were used to evaluate the robustness of the fusion model. To simulate the HrMSI, the SRF for the blue to SWIR2 bands of the Landsat 8 were used \cite{barsi2014spectral}. To verify the stability of the model, different GSD ratios and number bands were used to simulate the LrHSI and HrMSI. For the Pavia University and Indian Pines data, the GSD ratio was set to 4; it was set to 8 for the Washington DC and University of Houston imagery. The blue–green–red bands of the Landsat 8 SRF were used for Pavia University and University of Houston data, and the blue to SWIR2 part of the Landsat 8 SRF was used for Indian Pines and Washington DC. Table \ref{tab:table1} summarizes the simulated parameters for all of the datasets used in this experiment.

To evaluate the performances of different fusion methods, simulation experiments are used in this experiment. Simulation experiments refers to that the spatial and spectral down-sampling are implemented on the original HrHSI, and this one is the truth target image to evaluate the performance of the estimated HrHSI. Five different quality measures were used to evaluate the performance of the fusion results: the root mean square error (RMSE), peak SNR (PSNR), spectral angle mapper (SAM), relative global dimension error (ERGAS), and mean relative absolute error (MRAE) \cite{yokoya2017hyperspectral, arad2018ntire}. Of these measures, the RMSE, MRAE, and SAM were used in the visual evaluation, and the PSNR, SAM, and ERGAS were used in the quantitative evaluation.

The proposed model was trained using an Adam optimizer \cite{kingma2014adam} with the default parameters ${\beta _1} = 0.9$, ${\beta _2} = 0.999$ and $\varepsilon  = {10^{ - 8}}$; the initial learning rate was set to $5 \times {10^{ - 3}}$. Learning rate schedules seek to adjust the learning rate during training by reducing the learning rate according to a linear decay. After a total of 10000 epochs, the learning rate drops to 0. In our experiment, the number of input images was one; therefore, 1 epoch was equal to 1 iteration and the batch size was also 1. The Pytorch deep learning framework was used to train the proposed model \cite{paszke2019pytorch}. The training environment consisted of an Intel i7-6850K CPU, 128-GB RAM, and 4$\times$NVIDIA TITAN Xp 12G GPU.

\subsection{Parameters Discussion}

\begin{figure*}[!t]
	  \centering
		\subfigure{
			\includegraphics[width=0.9\textwidth]{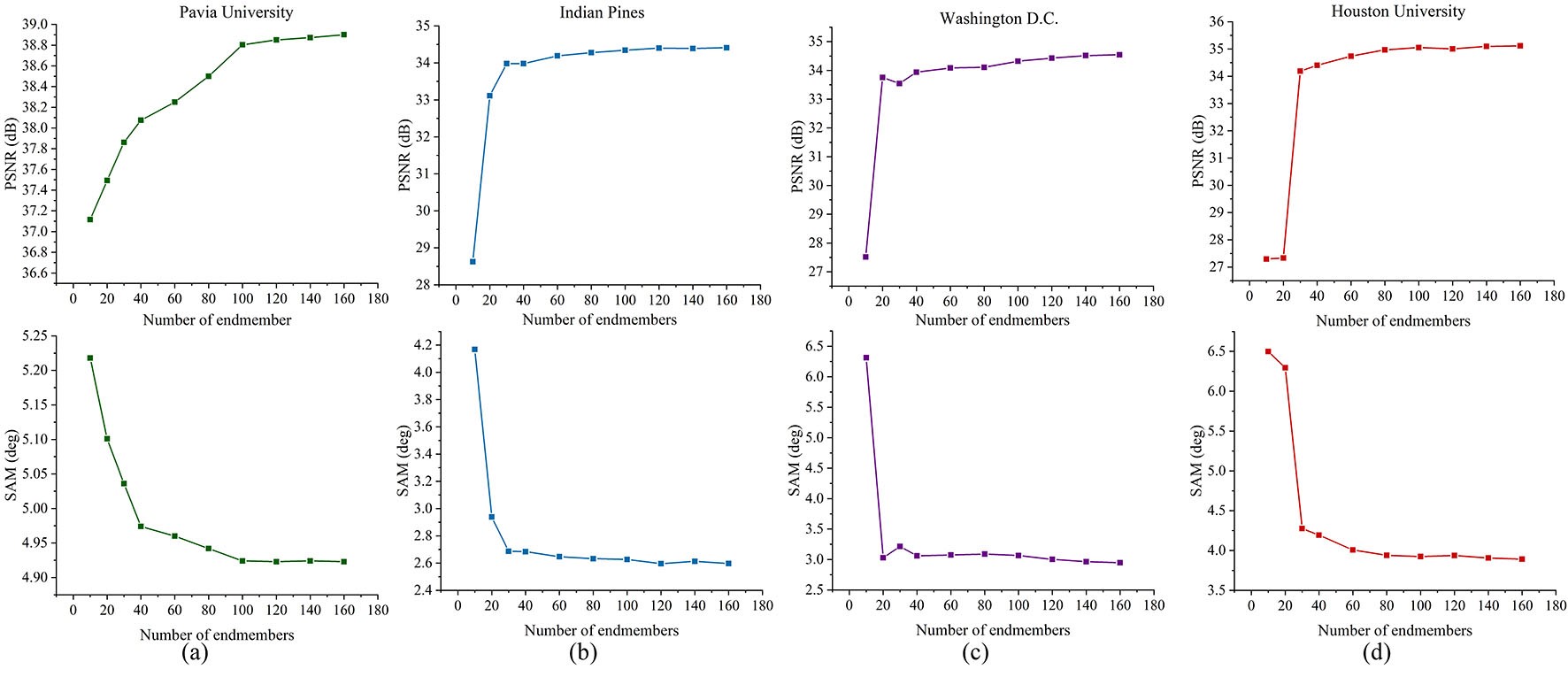}
		}
        \caption{The effect of endmember number on the fusion quality for different datasets. (a) Pavia University, (b) Indian Pines, (c) Washington DC, (d) University of Houston.}
\label{fig:fig8}
\end{figure*}

In the proposed method, the parameters $\alpha$, $\beta$, $\gamma$, $\mu$ and $\nu$ in the loss function Eqs. \ref{equation15} and \ref{equation18} need to be set. The parameters $\alpha$, $\beta$ and $\gamma$ are used to balance the weights of the different reconstruction errors. The parameters $\mu$ and $\nu$ are the trade-off parameters for the sum-to-one loss and sparsity loss.

Due to the fact that, in our method, the HSI–MSI fusion is driven by the autoencoder network, the reconstruction errors are the main factor that influence this process. Therefore, firstly, we set the parameters $\gamma$, $\mu$ and $\nu$ to a fixed value of 1 to evaluate the effect of using different values of the parameters $\alpha$ and $\beta$. In this part of the experiment, we used the Pavia University data to investigate the performance. Fig. \ref{fig:fig4} shows the PSNR of the target image $\mathbf{X}$ for different values of the trade-off parameters $\alpha$ and $\beta$. It can be seen that both parameters have a considerable effect on the fusion quality with the results being more sensitive to $\beta$ than to $\alpha$. This is because $\beta$ controls the weight of the reconstructed HrMSI, and therefore affects the spatial quality of the fused image. $\alpha$ also has an important effect on the quality of the fused target image. It is the weight of the $\mathbf{\tilde Z}_b$ reconstruction error and the process of reconstructing image $\mathbf{\tilde Z}_b$ is constrained by both the LrHSI and HrMSI autoencoder networks. Overall, the target image has a high PSNR when the two parameters $\alpha$ and $\beta$ are equal: these PSNR values have been marked in Fig. \ref{fig:fig4}. Clearly the PSNR increases as these two parameters increase and tends to be smoother when $\alpha=\beta=1$. The best result is achieved for $\alpha=\beta=10$. Therefore, we set $\alpha$ and $\beta$ to 10 in the later experiments.

In the second experiment, we tested the effect of parameter $\gamma$, which controls the weight for the reconstruction of the LrMSI images. Fig. \ref{fig:fig5} shows the experiment results. It can be seen that $\gamma$ has only a slight effect on the fusion quality. Since the reconstruction accuracy is higher and more stable when $\gamma=100$, we set $\gamma$ to 100 in the later experiments.

In the third experiment, the effects of $\mu$ and $\nu$ were investigated and the results are shown in Fig. \ref{fig:fig6}: $\mu$ is the weight of the sum-to-one loss and $\nu$ is the weight of the sparsity loss. It can clearly be seen that the performance is sensitive to $\mu$ as this parameter affects the precision of the fusion reconstruction. In contrast, the performance is not sensitive to $\nu$ since we set the number of endmembers $p$ to be 100. This is because the larger the number of endmembers, with the sum-to-one property, the more likely the endmembers are to be sparse. More details about the experiment carried out using different numbers of endmembers will be given in part \ref{numendmember}. Accordingly, we set $\mu$ and $\nu$ to 0.001 in the subsequent experiments.

To investigate the essentiality of the proposed network, as shown in Fig. \ref{fig:fig6-2}, ablation study was implemented in the case of the hyperparameters setting mentioned above. In this experiment, we can clearly see the performances when missing certain parts of the network or losses. As shown in Fig. \ref{fig:fig6-2}, $\not\subset$ means removing certain parts of the network or losses. It can be seen that the performances suddenly drop when removing $\mathbf{\tilde{Z}_b}$. Compared to this branch, removing one of the other branches interacts only with less effect. This indicates the branch of $\mathbf{\tilde{Z}_b}$ strongly ablation affects the overall fusion performance. Moreover, the branch $\mathbf{\tilde{Z}_b}$ means that the core advantage of the learnable PSF layer play an important role in improving the fusion performance.

\subsection{Constraint Function for the Abundance}\label{constraintfuc}

\begin{figure*}[!t]
	  \centering
		\subfigure{
			\includegraphics[width=0.9\textwidth]{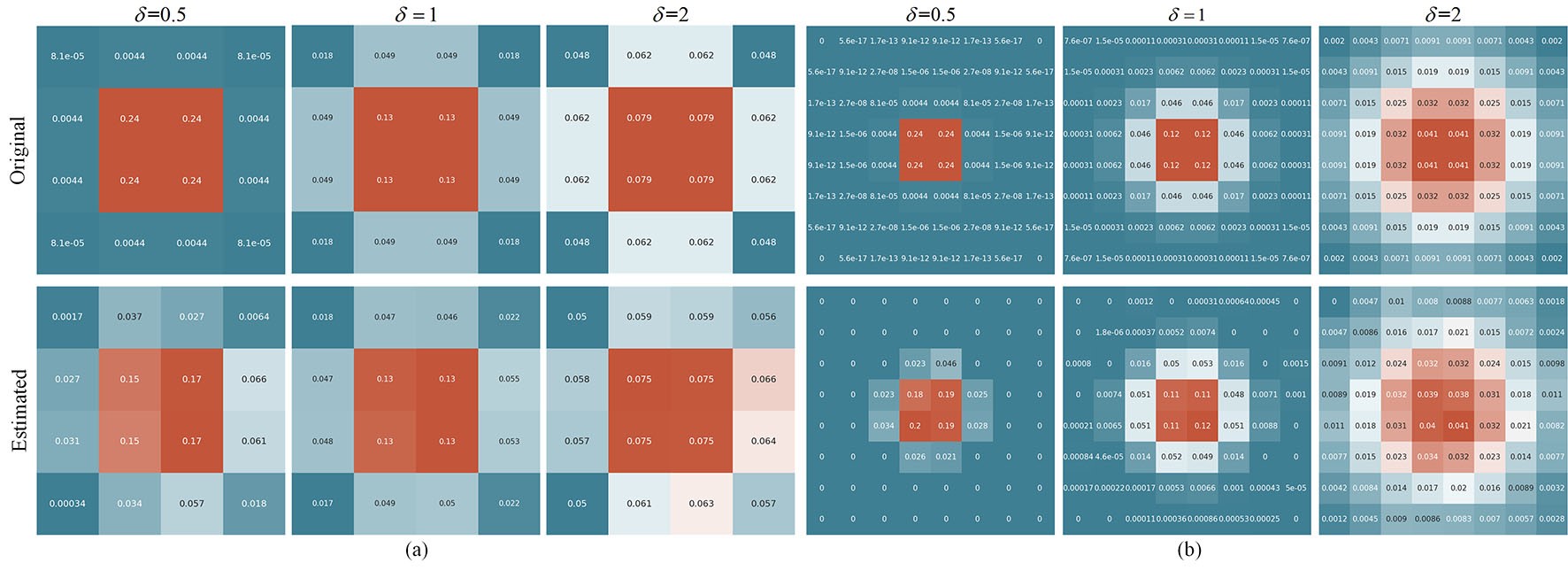}
		}
        \caption{The visualization of original and estimated PSF kernel. (a) Pavia University, (b) Washington DC.}
\label{fig:fig9}
\end{figure*}

\begin{figure}[!t]
    \centering
        \subfigure{
            \includegraphics[width=0.45\textwidth]{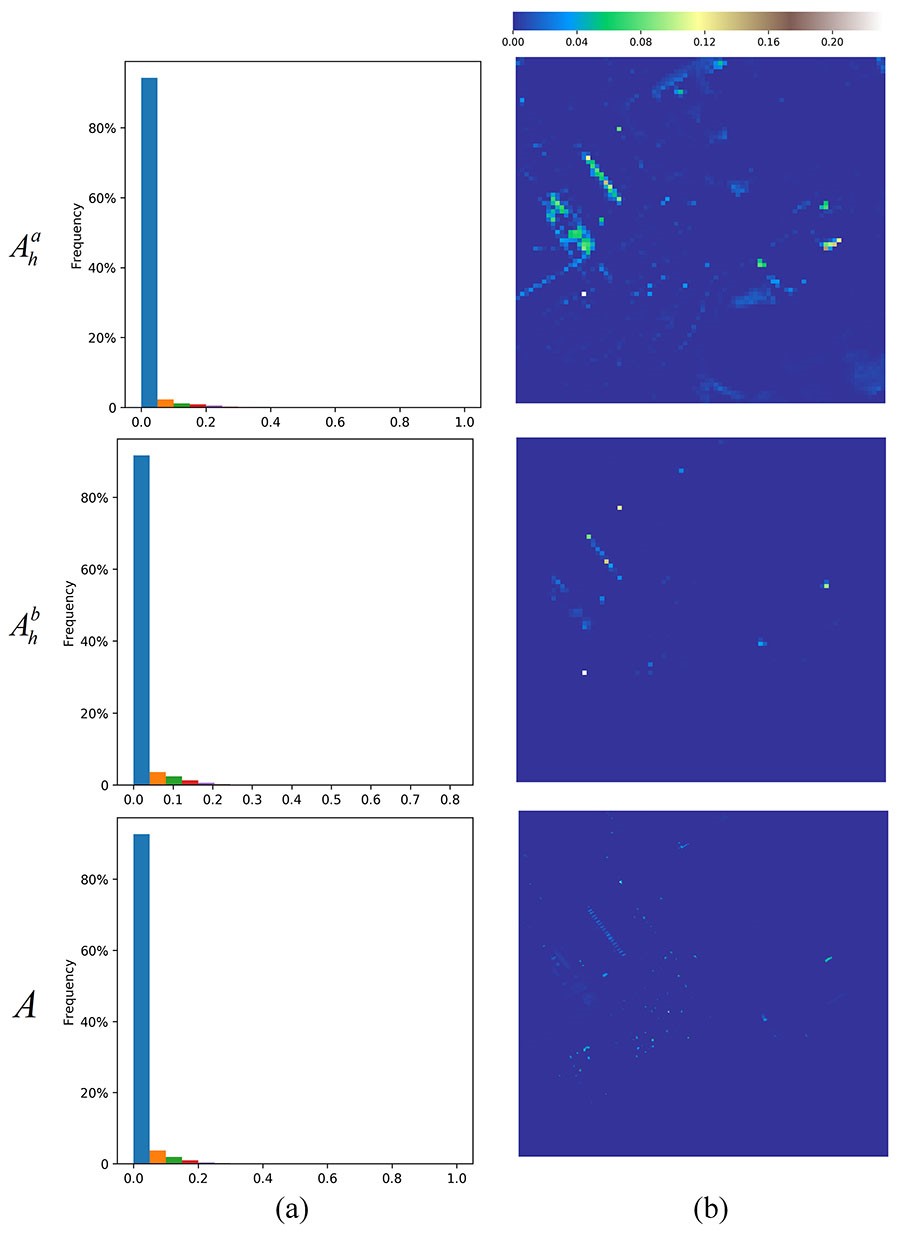}
        }
    \caption{The characters of learned abundances of three estimated abundances (Pavia University as example). (a) histograms, (b) the heatmaps of errors for the sum of abundance.}
    \label{fig:fig10}
\end{figure}

Eq. (\ref{equation16}) indicates that the sum-to-one property is restricted in the loss function; furthermore, a clamp function is used before the abundance to produce non-negative abundances. \cite{hong2019wu} reported that combing a softmax function before the abundances can also perfectly restrict the property of sum-to-one and nonnegative. However, in contrast to the clamp function, we found that using the softmax function leads to a lower reconstruction accuracy. Fig. \ref{fig:fig7a} shows the fusion accuracy obtained when using different constraint functions for the abundances. It can be observed that the fusion accuracy obviously improves when the clamp function is used. Although the softmax function has a smoother curve, the convergence accuracy and convergence speed are slightly lower.

Fig. \ref{fig:fig7b} can explain the results of this experiment: the figures shows the curves for the clamp and softmax functions. It can be seen that, due to the characteristics of the softmax function, it does not converge uniformly, meaning that different points converge at different rates and may converge arbitrarily slowly. This causes the gradient to be smaller for points with smaller values, which is equivalent to the gradient vanishing. In contrast, the gradient of the clamp function is updated faster in the range [0, 1]. 

\subsection{Number of Endmembers}\label{numendmember}

In general, the reconstruction accuracy improves as the number of endmembers, $p$, increases. Fig. \ref{fig:fig8} shows the changes in the PSNR and SAM with the number of endmembers for the four experimental datasets. To fully explore the effect of the hyperparameter $p$, the experiment was repeated three times for each value of $p$ using the same environment. The average values of the PSNR and SAM are shown in Fig. \ref{fig:fig8}. In our model, the number of endmembers $p$ represents the number of feature size of abundance and also represents the kernel size of the shared convolution layer. Therefore, a larger number of endmembers allows the model to be more representative. Although the number of endmembers is assumed to be equal to the number of pure spectral bases in the linear unmixing, the number of endmembers can also be larger than the actual number of pure bases because the convolution weight matrix $\mathbf{E}$ can contain mixed material \cite{yokoya2011coupled}. In addition, the convergence accuracy depends on the image complexity. In the experiments using the Indian Pines and Houston University data, the reconstruction accuracy began to converge at $p=30$, and fast convergence also occurred with the Washington DC data. For the Pavia University data, although the convergence was slow, the results were acceptable for smaller values of $p$. For convenience, we set $p=100$ in all cases when exploring the performance of the proposed model.

\subsection{Learned PSF Kernel}\label{learnedpsf}

The LrHSI was simulated using a PSF with a filter corresponding to the Gaussian function equal to the ratio of the GSD resolutions. Using the Pavia University and Washington DC data, we evaluated the learned kernels under different conditions. For each dataset, the standard deviations of the Gaussian kernel were set to 0.5, 1, and 2.

The original PSF kernels and the estimated ones are shown in Fig. \ref{fig:fig9}. Since the GSD ratios for the Pavia University and Washington DC data were 4 and 8, respectively, the kernel sizes for the two datasets were also 4 and 8, respectively. For Fig. \ref{fig:fig9}, it can be seen that the estimated kernels are similar to the original ones. A large standard deviation produces a smoother kernel and a large weighted combination of contributions from the local pixel. From the above experimental results, it can be seen that the proposed method is highly suitable for estimating arbitrary PSFs for different datasets.

\subsection{Estimated Abundances}

\begin{figure}[!t]
    \centering
        \subfigure{
            \includegraphics[width=0.39\textwidth]{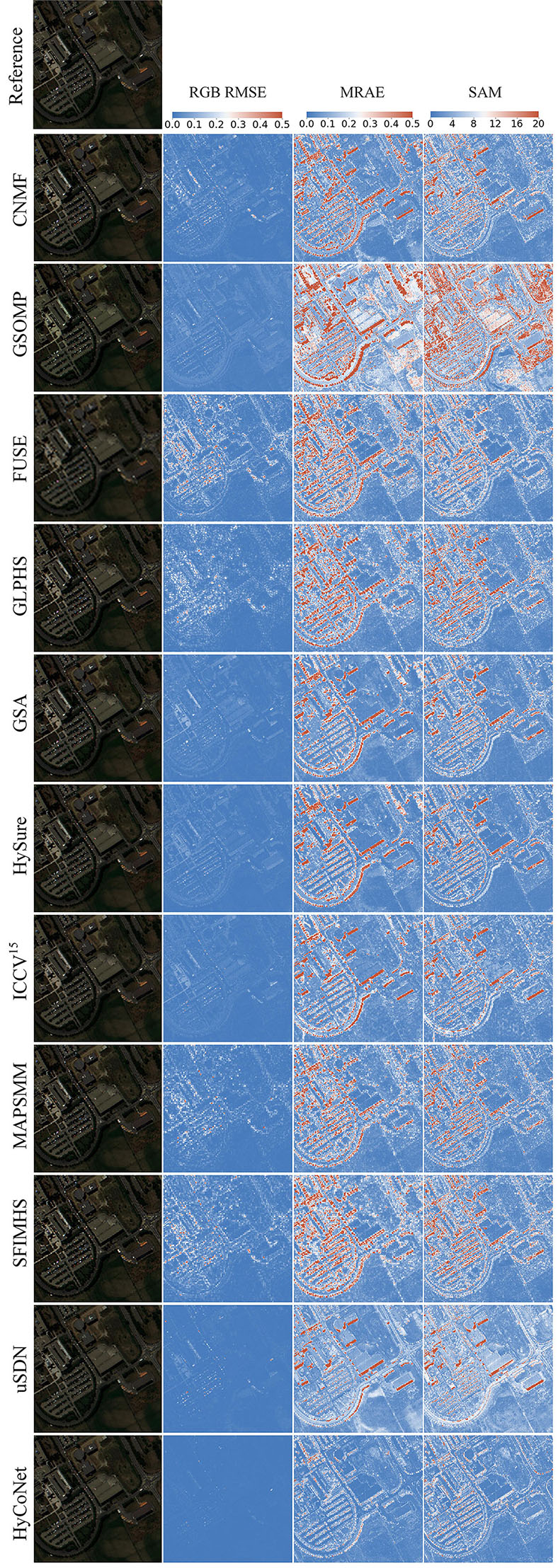}
        }
    \caption{The visualization of Pavia University fusion results. The first column is the color-composite of fusion results; the second column is RMSE error of color-composite image; the third column is the MARE error of HSI cube; the fourth column is the SAM error of HSI cube.}
    \label{fig:fig11}
\end{figure}

\begin{figure}[!t]
    \centering
        \subfigure{
            \includegraphics[width=0.39\textwidth]{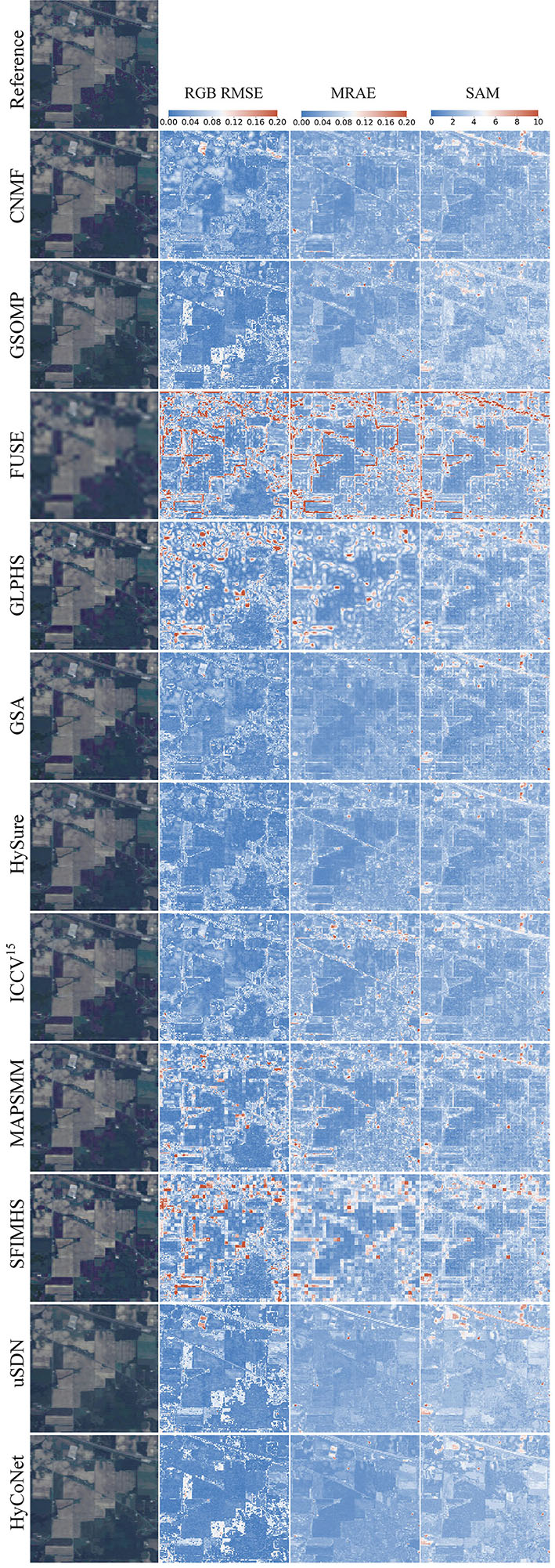}
        }
    \caption{The visualization of Indian Pines fusion results. The first column is the color-composite of fusion results; the second column is RMSE error of color-composite image; the third column is the MARE error of HSI cube; the fourth column is the SAM error of HSI cube.}
    \label{fig:fig12}
\end{figure}

\begin{figure}[!t]
    \centering
        \subfigure{
            \includegraphics[width=0.39\textwidth]{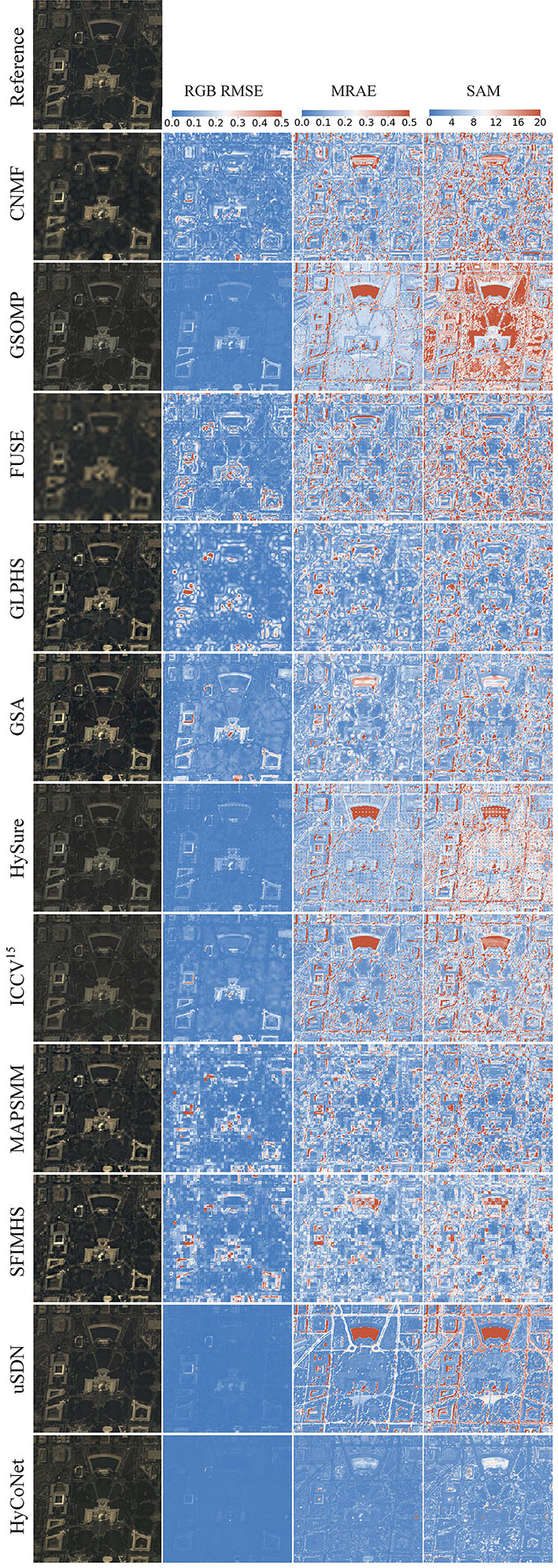}
        }
    \caption{The visualization of Washington DC fusion results. The first column is the color-composite of fusion results; the second column is RMSE error of color-composite image; the third column is the MARE error of HSI cube; the fourth column is the SAM error of HSI cube.}
    \label{fig:fig13}
\end{figure}

\begin{figure}[!t]
    \centering
        \subfigure{
            \includegraphics[width=0.39\textwidth]{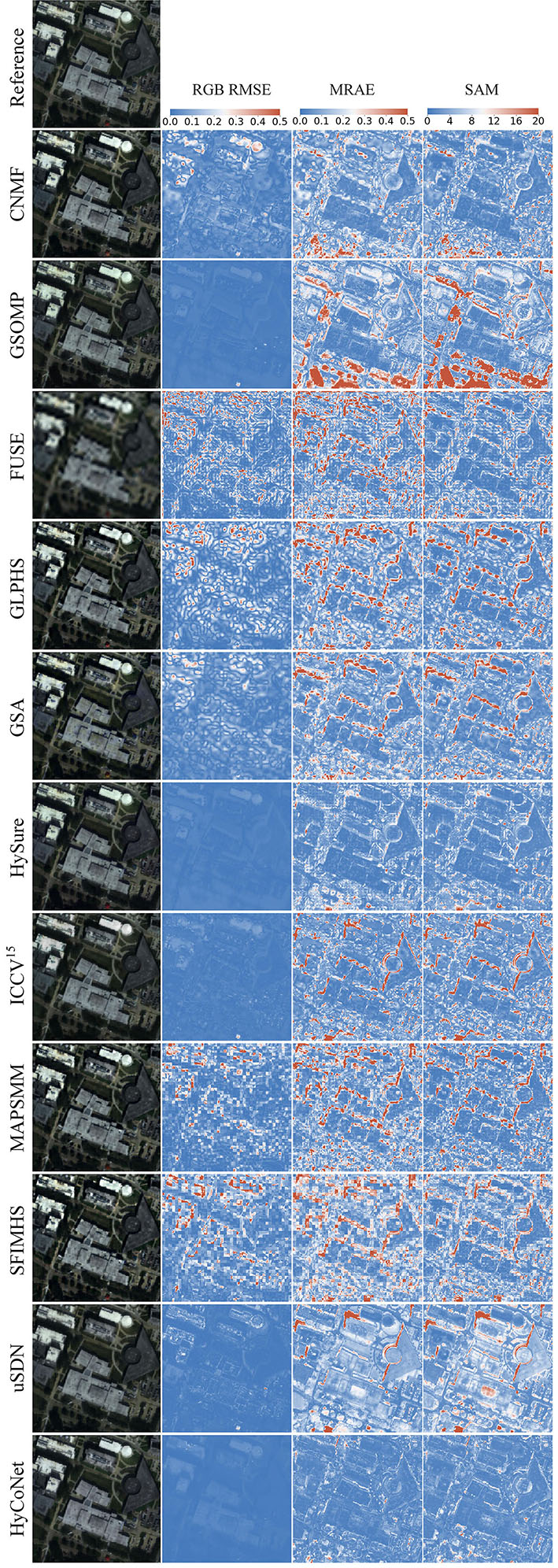}
        }
    \caption{The visualization of Houston University fusion results. The first column is the color-composite of fusion results; the second column is RMSE error of color-composite image; the third column is the MARE error of HSI cube; the fourth column is the SAM error of HSI cube.}
    \label{fig:fig14}
\end{figure}

In our network, the abundance is constrained by the sparse and sum-to-one characteristics of the loss function. Therefore, we next used the Pavia University data to explore the three estimated abundances corresponding to $\mathbf{A}_h^a$, $\mathbf{A}_h^b$ and $\mathbf{A}$. The histograms of these abundances are shown in Fig. \ref{fig:fig10}(a). The figure clearly shows that the estimated abundances are sparse. Fig. \ref{fig:fig10}(b) shows heatmaps of the errors relative to the sum of the abundances. The heatmap for $\mathbf{A}_h^a$ shows that some of the areas at the edges did not completely satisfy the sum-to-one constraint. The reason for this is that although the sum-to-one character is constrained by the loss function, it is possible that errors remain for some pixels. In contrast, the results for $\mathbf{A}_h^b$ and $\mathbf{A}$ are much better because they are constrained simultaneously by the LrHSI and HrMSI autoencoders. The abundances in our network, therefore, mostly do have sparse and sum-to-one characteristics.

\subsection{Comparison with the State of the Art}\label{comparesoa}

\begin{table*}[!t]
\scriptsize
\centering
\caption{Quantitative performance comparison with the different algorithms on the Pavia University data. The best one is shown in bold.}
\resizebox{\textwidth}{!}{


    \begin{tabular}{ccccccccccccc}
    \toprule
    \toprule
          &       & CNMF  & GSOMP & FUSE  & GLPHS & GSA   & HySure & ICCV15 & MAPSMM & SFIMHS & uSDN  & HyCoNet \\
    \midrule
    \multirow{3}[2]{*}{$\sigma=0.5$} & mSAM  & 4.8715  & 10.5319  & 4.7626  & 5.4793  & 4.1585  & 4.0756  & 4.4325  & 5.1705  & 5.4588  & 5.5614  & \textbf{3.4107 } \\
          & mPSNR & 31.8724  & 29.0635  & 27.7470  & 30.4492  & 34.3943  & 34.5460  & 34.7154  & 30.8779  & 23.2267  & 33.6010  & \textbf{38.7647 } \\
          & ERGAS & 4.3905  & 8.7599  & 6.4135  & 4.6189  & 3.1379  & 3.3403  & 3.2346  & 4.4652  & 25.2178  & 4.2374  & \textbf{2.8285 } \\
    \midrule
    \multirow{3}[2]{*}{$\sigma=1$} & mSAM  & 4.4029  & 9.0941  & 4.4667  & 4.7169  & 3.4810  & 3.5015  & 4.0694  & 4.0270  & 4.8534  & 5.3427  & \textbf{3.4002 } \\
          & mPSNR & 34.8032  & 31.0354  & 28.1708  & 32.9595  & 37.5222  & 36.9652  & 36.2953  & 34.9411  & 24.9070  & 33.1369  & \textbf{38.9132 } \\
          & ERGAS & 3.5658  & 9.2775  & 6.1141  & 3.7639  & \textbf{2.6955 } & 2.8068  & 2.8824  & 3.2969  & 20.8983  & 4.0075  & 2.7435  \\
    \midrule
    \multirow{3}[2]{*}{$\sigma=2$} & mSAM  & 3.7239  & 8.4744  & 4.4371  & 4.3608  & 3.4896  & 3.6169  & 4.0755  & 4.0317  & 4.5768  & 5.8754  & \textbf{3.4286 } \\
          & mPSNR & 36.0128  & 32.0907  & 28.2257  & 34.6668  & 38.4449  & 37.5251  & 36.7325  & 37.4392  & 27.3538  & 32.9005  & \textbf{38.5481 } \\
          & ERGAS & 3.1866  & 8.4769  & 6.0774  & 3.4117  & \textbf{2.6507 } & 2.9766  & 2.8770  & 3.1937  & 16.5452  & 4.1957  & 2.8722  \\
    \bottomrule
    \bottomrule
    \end{tabular}%

}
\label{tab:table2}
\end{table*}

\begin{table*}[!t]
\scriptsize
\centering
\caption{Quantitative performance comparison with the different algorithms on the Indian Pines data. The best one is shown in bold.}
\resizebox{\textwidth}{!}{
    \begin{tabular}{ccccccccccccc}
    \toprule
    \toprule
          &       & CNMF  & GSOMP & FUSE  & GLPHS & GSA   & HySure & ICCV15 & MAPSMM & SFIMHS & uSDN  & HyCoNet \\
    \midrule
    \multirow{3}[2]{*}{$\sigma=0.5$} & mSAM  & 2.4152  & 2.9762  & 3.4716  & 2.7551  & 2.4707  & 2.4194  & 2.4718  & 2.5734  & 2.9597  & 3.0325  & \textbf{2.3211 } \\
          & mPSNR & 32.4564  & 32.2554  & 26.7902  & 30.3093  & 33.6348  & 32.9597  & 31.3817  & 31.1324  & 28.9370  & 32.9082  & \textbf{34.0320 } \\
          & ERGAS & 1.4630  & 1.5615  & 2.6297  & 1.7735  & 1.3428  & 1.3927  & 1.6685  & 1.6691  & 2.0516  & 1.5496  & \textbf{1.3236 } \\
    \midrule
    \multirow{3}[2]{*}{$\sigma=1$} & mSAM  & 2.2617  & 2.8014  & 3.3703  & 2.3955  & 2.2692  & 2.2854  & 2.3812  & 2.3034  & 2.5387  & 2.9288  & \textbf{2.2447 } \\
          & mPSNR & 33.4572  & 32.4425  & 27.2244  & 32.9145  & 34.2168  & 33.6857  & 32.0225  & 33.5899  & 31.5298  & 31.1590  & \textbf{34.3232 } \\
          & ERGAS & 1.3687  & 1.5447  & 2.5078  & 1.3870  & 1.2040  & 1.3392  & 1.5949  & 1.3654  & 1.5670  & 1.6710  & \textbf{1.1946 } \\
    \midrule
    \multirow{3}[2]{*}{$\sigma=2$} & mSAM  & 2.2378  & 2.6955  & 3.3703  & 2.2745  & 2.2363  & 2.2939  & 2.3766  & 2.2395  & 2.3768  & 2.6423  & \textbf{2.2022 } \\
          & mPSNR & 33.7191  & 32.8198  & 27.2985  & 34.4669  & 34.6838  & 33.5168  & 32.0668  & 34.5392  & 33.1865  & 31.9779  & \textbf{34.7950 } \\
          & ERGAS & 1.3535  & 1.4978  & 2.4895  & 1.2376  & 1.1560  & 1.3768  & 1.5928  & 1.2865  & 1.3548  & 1.6187  & \textbf{1.1203 } \\
    \bottomrule
    \bottomrule
    \end{tabular}%

}
\label{tab:table3}
\end{table*}

\begin{table*}[!t]
\scriptsize
\centering
\caption{Quantitative performance comparison with the different algorithms on the Washington DC data. The best one is shown in bold.}
\resizebox{\textwidth}{!}{

    \begin{tabular}{ccccccccccccc}
    \toprule
    \toprule
          &       & CNMF  & GSOMP & FUSE  & GLPHS & GSA   & HySure & ICCV15 & MAPSMM & SFIMHS & uSDN  & HyCoNet \\
    \midrule
    \multirow{3}[2]{*}{$\sigma=0.5$} & mSAM  & 8.8273  & 13.1330  & 8.6505  & 6.9509  & 7.1772  & 9.9828  & 8.8196  & 7.3361  & 7.2102  & 7.0720  & \textbf{3.1984 } \\
          & mPSNR & 25.8781  & 25.3023  & 24.9366  & 27.5713  & 27.3163  & 26.4899  & 27.5002  & 26.9298  & 25.2003  & 33.0413  & \textbf{34.8730 } \\
          & ERGAS & 2.8031  & 3.4564  & 3.0064  & 2.2598  & 2.2676  & 2.7298  & 2.6514  & 2.4260  & 3.4973  & 1.6220  & \textbf{1.4334 } \\
    \midrule
    \multirow{3}[2]{*}{$\sigma=1$} & mSAM  & 7.2022  & 10.9944  & 8.2811  & 5.8102  & 5.8450  & 7.7087  & 8.0205  & 5.9814  & 6.0686  & 7.5995  & \textbf{3.2828 } \\
          & mPSNR & 27.8536  & 27.9343  & 25.7046  & 29.3968  & 30.0065  & 28.7836  & 29.9389  & 28.8842  & 26.9397  & 31.1618  & \textbf{34.9646 } \\
          & ERGAS & 2.2999  & 3.1083  & 2.8820  & 1.8748  & 1.7830  & 2.3315  & 2.2540  & 1.9589  & 36.0251  & 2.1821  & \textbf{1.3959 } \\
    \midrule
    \multirow{3}[2]{*}{$\sigma=2$} & mSAM  & 5.8034  & 9.6210  & 7.9871  & 4.1380  & 4.8727  & 6.5962  & 7.3153  & 3.7289  & 4.4360  & 4.8462  & \textbf{3.4581 } \\
          & mPSNR & 32.6577  & 28.9048  & 26.2474  & 33.4368  & 35.6293  & 29.8371  & 31.1686  & 33.7772  & 30.8051  & 32.8157  & \textbf{35.8561 } \\
          & ERGAS & 1.6080  & 2.6798  & 2.7159  & 1.3183  & 1.3323  & 2.0079  & 1.9705  & \textbf{1.2007 } & 2.5060  & 1.4754  & 1.4778  \\
    \bottomrule
    \bottomrule
    \end{tabular}%
}
\label{tab:table4}
\end{table*}

\begin{table*}[!t]
\scriptsize
\centering
\caption{Quantitative performance comparison with the different algorithms on the University of Houston data. The best one is shown in bold.}
\resizebox{\textwidth}{!}{

    \begin{tabular}{ccccccccccccc}
    \toprule
    \toprule
          &       & CNMF  & GSOMP & FUSE  & GLPHS & GSA   & HySure & ICCV15 & MAPSMM & SFIMHS & uSDN  & HyCoNet \\
    \midrule
    \multirow{3}[2]{*}{$\sigma=0.5$} & mSAM  & 4.1054  & 7.8904  & 4.5623  & 5.3563  & 5.0434  & 3.2919  & 4.7705  & 5.5041  & 5.3316  & 5.7001  & \textbf{2.6070 } \\
          & mPSNR & 27.0969  & 28.5365  & 24.0123  & 25.7426  & 27.7226  & 33.0650  & 31.3685  & 25.2709  & 22.5418  & 29.1871  & \textbf{35.2361 } \\
          & ERGAS & 1.9525  & 3.0026  & 2.5708  & 2.0911  & 1.7382  & 1.3990  & 1.5260  & 2.2232  & 4.5349  & 2.1537  & \textbf{1.0251 } \\
    \midrule
    \multirow{3}[2]{*}{$\sigma=1$} & mSAM  & 3.2652  & 7.9230  & 4.3228  & 4.7508  & 4.3542  & 3.2576  & 4.7106  & 4.3736  & 4.8702  & 4.9899  & \textbf{2.6670 } \\
          & mPSNR & 29.5577  & 28.9063  & 24.4412  & 26.9179  & 29.1212  & 34.0103  & 31.8105  & 26.9350  & 23.7143  & 29.0370  & \textbf{35.1123 } \\
          & ERGAS & 1.4203  & 2.9949  & 2.4456  & 1.8281  & 1.5293  & 1.3078  & 1.4552  & 1.8481  & 3.3914  & 2.0462  & \textbf{1.0252 } \\
    \midrule
    \multirow{3}[2]{*}{$\sigma=2$} & mSAM  & 3.2239  & 7.6917  & 4.1386  & 3.6522  & 3.5930  & 3.0419  & 4.1594  & 3.0458  & 4.0967  & 5.0680  & \textbf{2.6976 } \\
          & mPSNR & 31.0872  & 30.0214  & 24.9240  & 29.6952  & 32.8512  & 35.4746  & 32.7251  & 31.2549  & 26.2060  & 29.4494  & \textbf{35.5182 } \\
          & ERGAS & 1.2696  & 2.8725  & 2.3121  & 1.3591  & 1.2708  & 1.1233  & 1.3302  & 1.2350  & 2.3154  & 1.9161  & \textbf{1.0379 } \\
    \bottomrule
    \bottomrule
    \end{tabular}%
}
\label{tab:table5}
\end{table*}

\begin{figure*}[!t]
    \centering
        \subfigure{
            \includegraphics[width=0.95\textwidth]{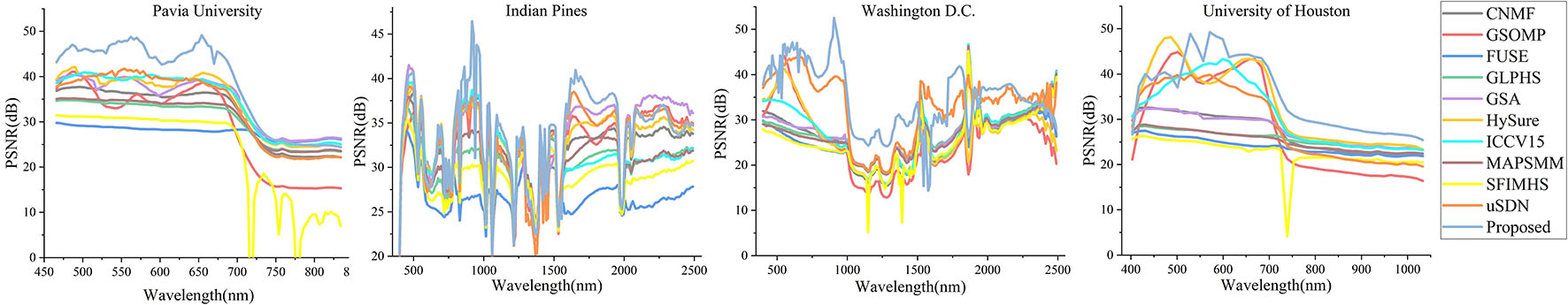}
        }
    \caption{The PSNR of different bands for comparison methods. (a) Pavia University; (b) Indian Pines; (c) Washington DC; (d) University of Houston.}
    \label{fig:fig15}
\end{figure*}

\subsubsection{Visual Performance}

Following the work of Yokoya \cite{yokoya2017hyperspectral}, in this study, a set of baseline methods were used for comparison. These included CNMF \cite{yokoya2011coupled}, GSOMP \cite{akhtar2014sparse}, FUSE \cite{wei2015fast}, GLPHS \cite{selva2015hyper}, GSA \cite{aiazzi2007improving}, HySure \cite{simoes2014convex}, Lanaras’s method (for convenience, we called it ICCV$^{15}$ because it was published in proceedings of the 2015 International Convention on Computer Vision) \cite{lanaras2015hyperspectral}, MAPSMM \cite{eismann2005hyperspectral}, SFIM-HS \cite{liu2000smoothing}, and uSDN \cite{qu2018unsupervised}. Because of the proposed fusion model is unsupervised algorithm and there is shortage of training samples required for supervised learning, only unsupervised fusion methods were used for fair comparison. This is also in line with the application requirements of real scenes. In addition, due to the fact that the proposed network is a fusion model where the SRF and PSF  are unknown and the only assumption is that prior information about the spectral coverage of the MSI is known, to conduct a fair comparison, the estimated SRFs obtained by the HySure SRF estimation method \cite{simoes2014convex} were used for the compared methods that require a SRF as input. These included CNMF, GSOMP, ICCV15, FUSE, MAPSMM and uSDN. The HySure SRF estimation method also only requires to know the spectral coverage of the HrMSI bands. Based on the GSD ratio of the simulated input image, a Gaussian kernel with a kernel size equal to the GSD ratio was used for all the methods which required a PSF kernel \cite{yokoya2017hyperspectral}; otherwise, the default source code settings were used.

Firstly, we used color-composite and heatmap images to visually evaluate the performance of the fusion results when $\sigma=0.5$. In Figs. \ref{fig:fig11}-\ref{fig:fig14}, the first column is the color-composite image (RGB image), the second column is the RMSE heatmap of the color-composite image, the third column is the MRAE heatmap, which can be considered to show the pixel-wise error for the reconstructed image cube, and the fourth column is the SAM error, which represents the spectral consistency of each pixel in the reconstructed image.

For most of the methods, the results for the color-composite images in the first column are good. However, for the heatmap in the second column, there are big differences between these methods. The RGB images for GSOMP, HySure, ICCV$^{15}$ and uSDN produce good results, indicating that these methods fully utilize the input HrMSI and the results retain more information about the HrMSI. Although the GSOMP method produces a good result for the RGB image, the heatmaps for MRAE and SAM are the poorest of those shown in Fig. \ref{fig:fig11}, \ref{fig:fig13} and \ref{fig:fig14}. A similar result was also reported by Yokoya et al \cite{yokoya2017hyperspectral}, who explained that the reason for this is that, in GSOMP, the high-resolution abundance is only estimated by the HrMSI and the sparsity prior. The errors for MAPSMM and FUSE show a block error distribution, which indicates that estimation of the results using patch-by-patch processing is unstable. For GSA and GLPHS, the errors have an inhomogeneous plaque block distribution. This is because the high-resolution image is obtained by sharpening the low-resolution image by adding spatial detail information and the cumulative error in the sharpened image can induce local irregular errors.

In the MRAE and SAM images, edge errors for objects are unavoidable for all the methods and this situation is particularly obvious in Figs. \ref{fig:fig11}, \ref{fig:fig13} and \ref{fig:fig14}. This can be explained by the fact that the mixing effect of low-resolution images makes it difficult to achieve better results in heterogeneous regions. Nonetheless, the results of our proposed method achieve the best visual results.

\subsubsection{Quantitative Performance}

We investigated the quantitative performance of all the compared methods, and the quality measures obtained are shown in Tables \ref{tab:table2}-\ref{tab:table5}. The mSAM is the mean of the SAM for all pixels and is used to evaluate the spectral consistency of the reconstructed HSI. The mPSNR is the mean PSNR of all the bands and is a measure of the spatial quality. The ERGAS is a global statistical measure used to evaluate the dimensionless global error for fused data.

From Tables \ref{tab:table2}-\ref{tab:table5}, it can be seen that our proposed method produces stable results for different datasets and different PSF deviations. However, the performance is unstable for all of the compared methods, especially for the Washington DC dataset. Due to the complexity of the objects in this imagery, most methods cannot handle the local relationship between the LrHSI and HSI when $\sigma=0.5$, this happened even to the methods that do not need prior knowledge of the PSF, e.g. GSA, SFIM-HS, GSOMP, HySure and MAP-SMM. For most methods, the performance improves as the standard deviation of the PSF increases. The results for the Indian Pines data are stable because the GSD for these data is the largest and the land objects are the simplest. The results for HySure, ICCV$^{15}$ and uSDN are stable except for the Washington DC data. Although an adaptive PSF is also implemented in the HySure model, the performance varies depending on the dataset used. Compared with the other methods, CNMF, FUSE, GSA, HySure and ICCV$^{15}$ have a better spectral consistency.

Fig. \ref{fig:fig15} shows the PSNR for the different bands of the HSI when $\sigma=0.5$, showing the reconstructed spatial quality for each band. It is clear that our proposed method significantly outperforms the other methods that were tested. For the Pavia University and University of Houston data, the results for HySure, ICCV$^{15}$ and GSOMP are good; however, again, all of the compared methods produce poor results for the Washington DC data. The results for the proposed method are not greatly affected by which dataset is used.

\section{Conclusion}

In this paper, we proposed a novel unsupervised deep learning method called HyCoNet to solve the HSI and MSI fusion problem for arbitrary PSFs and SRFs. Three coupled autoencoder nets were designed to extract spectral information from the LrHSI and spatial–contextual information from the HrMSI. Based on these autoencoder nets, the PSF was learned adaptively according to the correlation between the high- and low-resolution abundances, and the SRF was also learned by reconstruction of the autoencoder. Using the joint loss function, the proposed method can easily be implemented in an end-to-end training manner and provide a straightforward training strategy. The experiments that were performed indicated that the proposed method solved the HSI and MSI fusion problem without knowing the prior information of the PSF and SRF, and produced stable and robust fusion results for arbitrary PSFs and SRFs.


\section*{Acknowledgments}
The authors would like to thank Prof. David Landgrebe for providing the Indian Pines and Washington DC Mall data, Prof. Paolo Gamba for providing the Pavia University data, and the Hyperspectral Image Analysis group and the NSF Funded Center for Airborne Laser Mapping (NCALM) at the University of Houston for providing the CASI University of Houston data. The authors would like to express their appreciation to Dr. Naoto Yokoya for providing MATLAB codes for hyperspectral and multispectral data fusion toolbox and Dr. Ying Qu for providing the uSDN source code.

\bibliographystyle{IEEEtran}
\bibliography{HyCoNet_ref}

\begin{thebibliography}{10}
\providecommand{\url}[1]{#1}
\csname url@samestyle\endcsname
\providecommand{\newblock}{\relax}
\providecommand{\bibinfo}[2]{#2}
\providecommand{\BIBentrySTDinterwordspacing}{\spaceskip=0pt\relax}
\providecommand{\BIBentryALTinterwordstretchfactor}{4}
\providecommand{\BIBentryALTinterwordspacing}{\spaceskip=\fontdimen2\font plus
\BIBentryALTinterwordstretchfactor\fontdimen3\font minus
  \fontdimen4\font\relax}
\providecommand{\BIBforeignlanguage}[2]{{%
\expandafter\ifx\csname l@#1\endcsname\relax
\typeout{** WARNING: IEEEtran.bst: No hyphenation pattern has been}%
\typeout{** loaded for the language `#1'. Using the pattern for}%
\typeout{** the default language instead.}%
\else
\language=\csname l@#1\endcsname
\fi
#2}}
\providecommand{\BIBdecl}{\relax}
\BIBdecl

\bibitem{rasti2020feature}
B.~Rasti, D.~Hong, R.~Hang, P.~Ghamisi, X.~Kang, J.~Chanussot, and J.~A.
  Benediktsson, ``Feature extraction for hyperspectral imagery: The evolution
  from shallow to deep (overview and toolbox),'' \emph{IEEE Geosci. Remote
  Sens. Mag.}, 2020, dOI: 10.1109/MGRS.2020.2979764.

\bibitem{gao2014subspace}
L.~Gao, J.~Li, M.~Khodadadzadeh, A.~Plaza, B.~Zhang, Z.~He, and H.~Yan,
  ``Subspace-based support vector machines for hyperspectral image
  classification,'' \emph{IEEE Geosci. Remote Sens. Lett.}, vol.~12, no.~2, pp.
  349--353, 2014.

\bibitem{hong2020invariant}
D.~Hong, X.~Wu, P.~Ghamisi, J.~Chanussot, N.~Yokoya, and X.~X. Zhu, ``Invariant
  attribute profiles: A spatial-frequency joint feature extractor for
  hyperspectral image classification,'' \emph{IEEE Trans. Geosci. Remote
  Sens.}, vol.~58, no.~6, pp. 3791--3808, 2020.

\bibitem{cao2020hyperspectral}
X.~Cao, J.~Yao, Z.~Xu, and D.~Meng, ``Hyperspectral image classification with
  convolutional neural network and active learning,'' \emph{IEEE Trans. Geosci.
  Remote Sens.}, 2020, dOI:10.1109/TGRS.2020.2964627.

\bibitem{cao2020an}
X.~Cao, J.~Yao, X.~Fu, H.~Bi, and D.~Hong, ``An enhanced 3-dimensional discrete
  wavelet transform for hyperspectral image classification,'' \emph{IEEE
  Geosci. and Remote Sens. Lett.}, 2020, 10.1109/LGRS.2020.2990407.

\bibitem{guo2014weighted}
Q.~Guo, B.~Zhang, Q.~Ran, L.~Gao, J.~Li, and A.~Plaza, ``Weighted-rxd and
  linear filter-based rxd: Improving background statistics estimation for
  anomaly detection in hyperspectral imagery,'' \emph{IEEE J. Sel. Topics Appl.
  Earth Observ. Remote Sens.}, vol.~7, no.~6, pp. 2351--2366, 2014.

\bibitem{li2018real}
C.~Li, L.~Gao, Y.~Wu, B.~Zhang, J.~Plaza, and A.~Plaza, ``A real-time
  unsupervised background extraction-based target detection method for
  hyperspectral imagery,'' \emph{J. Real-Time Image Process.}, vol.~15, no.~3,
  pp. 597--615, 2018.

\bibitem{wu2019orsim}
X.~Wu, D.~Hong, J.~Tian, J.~Chanussot, W.~Li, and R.~Tao, ``Orsim detector: A
  novel object detection framework in optical remote sensing imagery using
  spatial-frequency channel features,'' \emph{IEEE Trans. Geosci. Remote
  Sens.}, vol.~57, no.~7, pp. 5146--5158, 2019.

\bibitem{wu2019fourier}
X.~Wu, D.~Hong, J.~Chanussot, Y.~Xu, R.~Tao, and Y.~Wang, ``Fourier-based
  rotation-invariant feature boosting: An efficient framework for geospatial
  object detection,'' \emph{IEEE Geosci. Remote Sens. Lett.}, vol.~17, no.~2,
  pp. 302--306, 2020.

\bibitem{he2016weighted}
W.~He, H.~Zhang, L.~Zhang, W.~Philips, and W.~Liao, ``Weighted sparse graph
  based dimensionality reduction for hyperspectral images,'' \emph{IEEE Geosci.
  Remote Sens. Lett.}, vol.~13, no.~5, pp. 686--690, 2016.

\bibitem{hong2017learning}
D.~Hong, N.~Yokoya, and X.~X. Zhu, ``Learning a robust local manifold
  representation for hyperspectral dimensionality reduction,'' \emph{IEEE J.
  Sel. Top. Appl. Earth Obs. Remote Sens.}, vol.~10, no.~6, pp. 2960--2975,
  2017.

\bibitem{xu2019superpixel}
H.~Xu, H.~Zhang, W.~He, and L.~Zhang, ``Superpixel-based spatial-spectral
  dimension reduction for hyperspectral imagery classification,''
  \emph{Neurocomputing}, vol. 360, pp. 138--150, 2019.

\bibitem{hong2019learning}
D.~Hong, N.~Yokoya, J.~Chanussot, J.~Xu, and X.~X. Zhu, ``Learning to propagate
  labels on graphs: An iterative multitask regression framework for
  semi-supervised hyperspectral dimensionality reduction,'' \emph{ISPRS J.
  Photogramm. Remote Sens.}, vol. 158, pp. 35--49, 2019.

\bibitem{xu2019nonlocal}
Y.~Xu, Z.~Wu, J.~Chanussot, P.~Comon, and Z.~Wei, ``Nonlocal coupled tensor cp
  decomposition for hyperspectral and multispectral image fusion,'' \emph{IEEE
  Trans. Geosci. Remote Sens.}, vol.~58, no.~1, pp. 348--362, 2019.

\bibitem{hu2019mima}
J.~Hu, D.~Hong, and X.~X. Zhu, ``{MIMA}: Mapper-induced manifold alignment for
  semi-supervised fusion of optical image and polarimetric sar data,''
  \emph{IEEE Trans. Geosci. Remote Sens.}, vol.~57, no.~11, pp. 9025--9040,
  2019.

\bibitem{hang2020classification}
R.~Hang, Z.~Li, P.~Ghamisi, D.~Hong, G.~Xia, and Q.~Liu, ``Classification of
  hyperspectral and lidar data using coupled cnns,'' \emph{IEEE Trans. Geosci.
  Remote Sens.}, vol.~58, no.~7, pp. 4939--4950, 2020.

\bibitem{tang2017integrating}
M.~Tang, L.~Gao, A.~Marinoni, P.~Gamba, and B.~Zhang, ``Integrating spatial
  information in the normalized p-linear algorithm for nonlinear hyperspectral
  unmixing,'' \emph{IEEE J. Sel. Top. Appl. Earth Obs. Remote Sens.}, vol.~11,
  no.~4, pp. 1179--1190, 2017.

\bibitem{hong2018sulora}
D.~Hong and X.~X. Zhu, ``S{UL}o{RA}: Subspace unmixing with low-rank attribute
  embedding for hyperspectral data analysis,'' \emph{IEEE J. Sel. Topics Signal
  Process.}, vol.~12, no.~6, pp. 1351--1363, 2018.

\bibitem{yao2019nonconvex}
J.~Yao, D.~Meng, Q.~Zhao, W.~Cao, and Z.~Xu, ``Nonconvex-sparsity and
  nonlocal-smoothness-based blind hyperspectral unmixing,'' \emph{IEEE Trans.
  Image Process.}, vol.~28, no.~6, pp. 2991--3006, 2019.

\bibitem{loncan2015hyperspectral}
L.~Loncan, L.~B. De~Almeida, J.~M. Bioucas-Dias, X.~Briottet, J.~Chanussot,
  N.~Dobigeon, S.~Fabre, W.~Liao, G.~A. Licciardi, M.~Simoes \emph{et~al.},
  ``Hyperspectral pansharpening: A review,'' \emph{IEEE Geosci. Remote Sens.
  Mag.}, vol.~3, no.~3, pp. 27--46, 2015.

\bibitem{hong2020learning}
D.~Hong, J.~Chanussot, N.~Yokoya, J.~Kang, and X.~X. Zhu, ``Learning shared
  cross-modality representation using multispectral-lidar and hyperspectral
  data,'' \emph{IEEE Geosci. Remote Sens. Lett.}, 2020,
  dOI:10.1109/LGRS.2019.2944599.

\bibitem{gomez2001wavelet}
R.~B. Gomez, A.~Jazaeri, and M.~Kafatos, ``Wavelet-based hyperspectral and
  multispectral image fusion,'' in \emph{Geo-Spatial Image and Data
  Exploitation II}, vol. 4383.\hskip 1em plus 0.5em minus 0.4em\relax
  International Society for Optics and Photonics, 2001, pp. 36--42.

\bibitem{zhang2007multi}
Y.~Zhang and M.~He, ``Multi-spectral and hyperspectral image fusion using 3-d
  wavelet transform,'' \emph{J. Electron.}, vol.~24, no.~2, pp. 218--224, 2007.

\bibitem{chen2014fusion}
Z.~Chen, H.~Pu, B.~Wang, and G.-M. Jiang, ``Fusion of hyperspectral and
  multispectral images: A novel framework based on generalization of
  pan-sharpening methods,'' \emph{IEEE Geosci. and Remote Sens. Lett.},
  vol.~11, no.~8, pp. 1418--1422, 2014.

\bibitem{aiazzi2007improving}
B.~Aiazzi, S.~Baronti, and M.~Selva, ``Improving component substitution
  pansharpening through multivariate regression of ms $+ $ pan data,''
  \emph{IEEE Trans. Geosci. Remote Sens.}, vol.~45, no.~10, pp. 3230--3239,
  2007.

\bibitem{liu2000smoothing}
J.~Liu, ``Smoothing filter-based intensity modulation: A spectral preserve
  image fusion technique for improving spatial details,'' \emph{Int. J. Remote
  Sens.}, vol.~21, no.~18, pp. 3461--3472, 2000.

\bibitem{eismann2005hyperspectral}
M.~T. Eismann and R.~C. Hardie, ``Hyperspectral resolution enhancement using
  high-resolution multispectral imagery with arbitrary response functions,''
  \emph{IEEE Trans. Geosci. Remote Sens.}, vol.~43, no.~3, pp. 455--465, 2005.

\bibitem{wei2015hyperspectral}
Q.~Wei, J.~Bioucas-Dias, N.~Dobigeon, and J.-Y. Tourneret, ``Hyperspectral and
  multispectral image fusion based on a sparse representation,'' \emph{IEEE
  Trans. Geosci. Remote Sens.}, vol.~53, no.~7, pp. 3658--3668, 2015.

\bibitem{simoes2014convex}
M.~Simoes, J.~Bioucas-Dias, L.~B. Almeida, and J.~Chanussot, ``A convex
  formulation for hyperspectral image superresolution via subspace-based
  regularization,'' \emph{IEEE Trans. Geosci. Remote Sens.}, vol.~53, no.~6,
  pp. 3373--3388, 2014.

\bibitem{akhtar2015bayesian}
N.~Akhtar, F.~Shafait, and A.~Mian, ``Bayesian sparse representation for
  hyperspectral image super resolution,'' in \emph{Proc. CVPR}.\hskip 1em plus
  0.5em minus 0.4em\relax IEEE, 2015, pp. 3631--3640.

\bibitem{kawakami2011high}
R.~Kawakami, Y.~Matsushita, J.~Wright, M.~Ben-Ezra, Y.-W. Tai, and K.~Ikeuchi,
  ``High-resolution hyperspectral imaging via matrix factorization,'' in
  \emph{Proc. CVPR}.\hskip 1em plus 0.5em minus 0.4em\relax IEEE, 2011, pp.
  2329--2336.

\bibitem{yokoya2011coupled}
N.~Yokoya, T.~Yairi, and A.~Iwasaki, ``Coupled nonnegative matrix factorization
  unmixing for hyperspectral and multispectral data fusion,'' \emph{IEEE Trans.
  Geosci. Remote Sens.}, vol.~50, no.~2, pp. 528--537, 2011.

\bibitem{lanaras2015hyperspectral}
C.~Lanaras, E.~Baltsavias, and K.~Schindler, ``Hyperspectral super-resolution
  by coupled spectral unmixing,'' in \emph{Proc. CVPR}.\hskip 1em plus 0.5em
  minus 0.4em\relax IEEE, 2015, pp. 3586--3594.

\bibitem{wycoff2013non}
E.~Wycoff, T.-H. Chan, K.~Jia, W.-K. Ma, and Y.~Ma, ``A non-negative sparse
  promoting algorithm for high resolution hyperspectral imaging,'' in
  \emph{Proc. ICASSP}.\hskip 1em plus 0.5em minus 0.4em\relax IEEE, 2013, pp.
  1409--1413.

\bibitem{akhtar2014sparse}
N.~Akhtar, F.~Shafait, and A.~Mian, ``Sparse spatio-spectral representation for
  hyperspectral image super-resolution,'' in \emph{Proc. ECCV}.\hskip 1em plus
  0.5em minus 0.4em\relax Springer, 2014, pp. 63--78.

\bibitem{yi2018hyperspectral}
C.~Yi, Y.-Q. Zhao, and J.~C.-W. Chan, ``Hyperspectral image super-resolution
  based on spatial and spectral correlation fusion,'' \emph{IEEE Trans. Geosci.
  Remote Sens.}, vol.~56, no.~7, pp. 4165--4177, 2018.

\bibitem{dian2017hyperspectral}
R.~Dian, L.~Fang, and S.~Li, ``Hyperspectral image super-resolution via
  non-local sparse tensor factorization,'' in \emph{Proc. CVPR}.\hskip 1em plus
  0.5em minus 0.4em\relax Springer, 2017, pp. 5344--5353.

\bibitem{long2017accurate}
Y.~Long, Y.~Gong, Z.~Xiao, and Q.~Liu, ``Accurate object localization in remote
  sensing images based on convolutional neural networks,'' \emph{IEEE Trans.
  Geosci. Remote Sens.}, vol.~55, no.~5, pp. 2486--2498, 2017.

\bibitem{wu2018msri}
X.~Wu, D.~Hong, P.~Ghamisi, W.~Li, and R.~Tao, ``Ms{R}i-{CCF}: Multi-scale and
  rotation-insensitive convolutional channel features for geospatial object
  detection,'' \emph{Remote Sens.}, vol.~10, no.~12, p. 1990, 2018.

\bibitem{jia2019collaborative}
S.~Jia, X.~Deng, J.~Zhu, M.~Xu, J.~Zhou, and X.~Jia, ``Collaborative
  representation-based multiscale superpixel fusion for hyperspectral image
  classification,'' \emph{IEEE Trans. Geosci. Remote Sens.}, vol.~57, no.~10,
  pp. 7770--7784, 2019.

\bibitem{liang2018material}
J.~Liang, J.~Zhou, L.~Tong, X.~Bai, and B.~Wang, ``Material based salient
  object detection from hyperspectral images,'' \emph{Pattern Recognit.},
  vol.~76, pp. 476--490, 2018.

\bibitem{wang2017deep}
C.~Wang, Y.~Liu, X.~Bai, W.~Tang, P.~Lei, and J.~Zhou, ``Deep residual
  convolutional neural network for hyperspectral image super-resolution,'' in
  \emph{Proc. ICIG}.\hskip 1em plus 0.5em minus 0.4em\relax Springer, 2017, pp.
  370--380.

\bibitem{han2018ssf}
X.-H. Han, B.~Shi, and Y.~Zheng, ``Ssf-cnn: Spatial and spectral fusion with
  cnn for hyperspectral image super-resolution,'' in \emph{Proc. ICIP}.\hskip
  1em plus 0.5em minus 0.4em\relax IEEE, 2018, pp. 2506--2510.

\bibitem{palsson2017multispectral}
F.~Palsson, J.~R. Sveinsson, and M.~O. Ulfarsson, ``Multispectral and
  hyperspectral image fusion using a 3-d-convolutional neural network,''
  \emph{IEEE Geosci. and Remote Sens. Lett.}, vol.~14, no.~5, pp. 639--643,
  2017.

\bibitem{dian2018deep}
R.~Dian, S.~Li, A.~Guo, and L.~Fang, ``Deep hyperspectral image sharpening,''
  \emph{IEEE Trans. Neural Netw. Learn. Syst.}, no.~99, pp. 1--11, 2018.

\bibitem{xie2019multispectral}
Q.~Xie, M.~Zhou, Q.~Zhao, D.~Meng, W.~Zuo, and Z.~Xu, ``Multispectral and
  hyperspectral image fusion by ms/hs fusion net,'' in \emph{Proc. CVPR}.\hskip
  1em plus 0.5em minus 0.4em\relax IEEE, 2019, pp. 1585--1594.

\bibitem{wang2019deep}
W.~Wang, W.~Zeng, Y.~Huang, X.~Ding, and J.~Paisley, ``Deep blind hyperspectral
  image fusion,'' in \emph{Proc. ICCV}, 2019, pp. 4150--4159.

\bibitem{han2019multi}
X.-H. Han, Y.~Zheng, and Y.-W. Chen, ``Multi-level and multi-scale spatial and
  spectral fusion cnn for hyperspectral image super-resolution,'' in
  \emph{Proc. ICCV workshops}, 2019, pp. 0--0.

\bibitem{qu2018unsupervised}
Y.~Qu, H.~Qi, and C.~Kwan, ``Unsupervised sparse dirichlet-net for
  hyperspectral image super-resolution,'' in \emph{Proc. CVPR}.\hskip 1em plus
  0.5em minus 0.4em\relax IEEE, 2018, pp. 2511--2520.

\bibitem{zhou2019integrated}
Y.~Zhou, A.~Rangarajan, and P.~D. Gader, ``An integrated approach to
  registration and fusion of hyperspectral and multispectral images,''
  \emph{IEEE Trans. Geosci. Remote Sens.}, 2019.

\bibitem{fu2019hyperspectral}
Y.~Fu, T.~Zhang, Y.~Zheng, D.~Zhang, and H.~Huang, ``Hyperspectral image
  super-resolution with optimized rgb guidance,'' in \emph{Proc. CVPR}, 2019,
  pp. 11\,661--11\,670.

\bibitem{hong2018augmented}
D.~Hong, N.~Yokoya, J.~Chanussot, and X.~X. Zhu, ``An augmented linear mixing
  model to address spectral variability for hyperspectral unmixing,''
  \emph{IEEE Trans. Image Process.}, vol.~28, no.~4, pp. 1923--1938, 2019.

\bibitem{kang2020learning}
J.~Kang, D.~Hong, J.~Liu, G.~Baier, N.~Yokoya, and B.~Demir, ``Learning
  convolutional sparse coding on complex domain for interferometric phase
  restoration,'' \emph{IEEE Trans. Neural Netw. Learn. Syst.}, 2020, dOI:
  10.1109/TNNLS.2020.2979546.

\bibitem{wang2017effect}
Q.~Wang and P.~M. Atkinson, ``The effect of the point spread function on
  sub-pixel mapping,'' \emph{Remote Sens Environ}, vol. 193, pp. 127--137,
  2017.

\bibitem{hong2015novel}
D.~Hong, W.~Liu, J.~Su, Z.~Pan, and G.~Wang, ``A novel hierarchical approach
  for multispectral palmprint recognition,'' \emph{Neurocomputing}, vol. 151,
  pp. 511--521, 2015.

\bibitem{yokoya2017hyperspectral}
N.~Yokoya, C.~Grohnfeldt, and J.~Chanussot, ``Hyperspectral and multispectral
  data fusion: A comparative review of the recent literature,'' \emph{IEEE
  Geosci. Remote Sens. Mag.}, vol.~5, no.~2, pp. 29--56, 2017.

\bibitem{le20182018}
B.~Le~Saux, N.~Yokoya, R.~Hansch, and S.~Prasad, ``2018 ieee grss data fusion
  contest: Multimodal land use classification [technical committees],''
  \emph{IEEE Geosci. Remote Sens. Mag.}, vol.~6, no.~1, pp. 52--54, 2018.

\bibitem{hong2019learnable}
D.~Hong, N.~Yokoya, N.~Ge, J.~Chanussot, and X.~X. Zhu, ``Learnable manifold
  alignment ({L}e{MA}): A semi-supervised cross-modality learning framework for
  land cover and land use classification,'' \emph{ISPRS J. Photogramm. Remote
  Sens.}, vol. 147, pp. 193--205, 2019.

\bibitem{barsi2014spectral}
J.~A. Barsi, K.~Lee, G.~Kvaran, B.~L. Markham, and J.~A. Pedelty, ``The
  spectral response of the landsat-8 operational land imager,'' \emph{Remote
  Sens.}, vol.~6, no.~10, pp. 10\,232--10\,251, 2014.

\bibitem{arad2018ntire}
B.~Arad, O.~Ben-Shahar, and R.~Timofte, ``Ntire 2018 challenge on spectral
  reconstruction from rgb images,'' in \emph{Proc. CVPR}.\hskip 1em plus 0.5em
  minus 0.4em\relax IEEE, 2018, pp. 929--938.

\bibitem{kingma2014adam}
D.~P. Kingma and J.~Ba, ``Adam: A method for stochastic optimization,'' in
  \emph{Proc. ICLR}, 2014.

\bibitem{paszke2019pytorch}
A.~Paszke, S.~Gross, F.~Massa, A.~Lerer, J.~Bradbury, G.~Chanan, T.~Killeen,
  Z.~Lin, N.~Gimelshein, L.~Antiga \emph{et~al.}, ``Pytorch: An imperative
  style, high-performance deep learning library,'' in \emph{Proc. NIPS}, 2019,
  pp. 8024--8035.

\bibitem{hong2019wu}
D.~Hong, J.~Chanussot, N.~Yokoya, U.~Heiden, W.~Heldens, and X.~X. Zhu,
  ``Wu-net: A weakly-supervised unmixing network for remotely sensed
  hyperspectral imagery,'' in \emph{Proc. IGARSS}.\hskip 1em plus 0.5em minus
  0.4em\relax IEEE, 2019, pp. 373--376.

\bibitem{wei2015fast}
Q.~Wei, N.~Dobigeon, and J.-Y. Tourneret, ``Fast fusion of multi-band images
  based on solving a sylvester equation,'' \emph{IEEE Trans. Image Process.},
  vol.~24, no.~11, pp. 4109--4121, 2015.

\bibitem{selva2015hyper}
M.~Selva, B.~Aiazzi, F.~Butera, L.~Chiarantini, and S.~Baronti,
  ``Hyper-sharpening: A first approach on sim-ga data,'' \emph{IEEE J. Sel.
  Topics Appl. Earth Observ. Remote Sens.}, vol.~8, no.~6, pp. 3008--3024,
  2015.

\end{thebibliography}

\vskip -2\baselineskip plus -1fil
\begin{IEEEbiography}[{\includegraphics[width=1in,height=1.25in,clip,keepaspectratio]{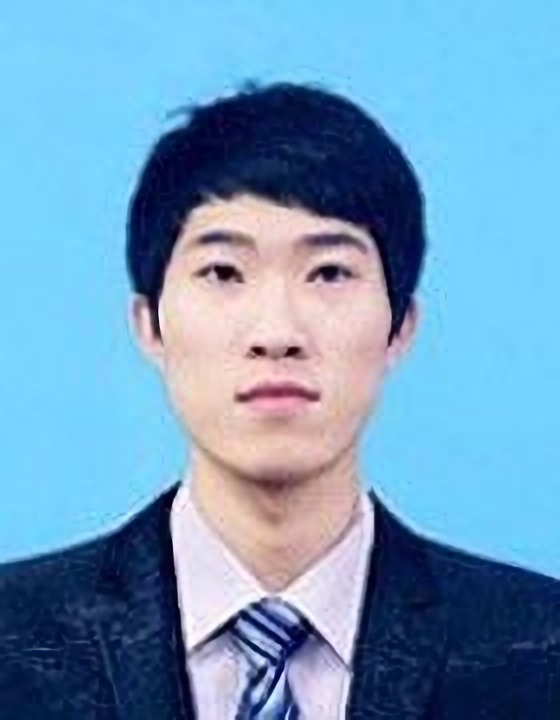}}]{Ke Zheng} received B.S. degree in geographic information system from Shandong Agricultural University, Taian, China, in 2008, and the M.S. and Ph.D. degrees in remote sensing from College of Geosciences and Surveying Engineering, China University of Mining and Technology (Beijing), Beijing, China, in 2016 and 2020, respectively.

Since 2020, he has been a Post-Doctoral Associate with the Key Laboratory of Digital Earth Science, Aerospace Information Research Institute, Chinese Academy of Science, Beijing, China. His research interests include image processing, machine learning, deep learning and their application in Earth Vision.
\end{IEEEbiography}

\vskip -2\baselineskip plus -1fil
\begin{IEEEbiography}[{\includegraphics[width=1in,height=1.25in,clip,keepaspectratio]{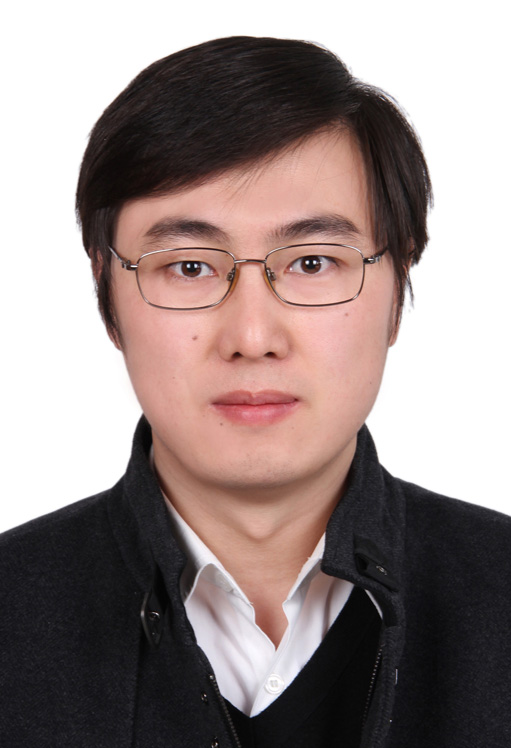}}]{Lianru Gao} (M'12-SM'18) received the B.S. degree in civil engineering from Tsinghua University, Beijing, China, in 2002, and the Ph.D. degree in cartography and geographic information system from Institute of Remote Sensing Applications, Chinese Academy of Sciences (CAS), Beijing, China, in 2007.

He is currently a Professor with the Key Laboratory of Digital Earth Science, Aerospace Information Research Institute, CAS. He also has been a visiting scholar at the University of Extremadura, Cáceres, Spain, in 2014, and at the Mississippi State University (MSU), Starkville, USA, in 2016. His research focuses on hyperspectral image processing and information extraction. In last ten years, he was the PI of 10 scientific research projects at national and ministerial levels, including projects by the National Natural Science Foundation of China (2010-2012, 2016-2019, 2018-2020), and by the Key Research Program of the CAS (2013-2015) et al. He has published more than 160 peer-reviewed papers, and there are more than 80 journal papers included by Science Citation Index (SCI). He was coauthor of an academic book entitled ``Hyperspectral Image Classification And Target Detection''. He obtained 28 National Invention Patents in China. He was awarded the Outstanding Science and Technology Achievement Prize of the CAS in 2016, and was supported by the China National Science Fund for Excellent Young Scholars in 2017, and won the Second Prize of The State Scientific and Technological Progress Award in 2018. He received the recognition of the Best Reviewers of the IEEE Journal of Selected Topics in Applied Earth Observations and Remote Sensing in 2015, and the Best Reviewers of the IEEE Transactions on Geoscience and Remote Sensing in 2017.
\end{IEEEbiography}

\begin{IEEEbiography}[{\includegraphics[width=1in,height=1.25in,clip,keepaspectratio]{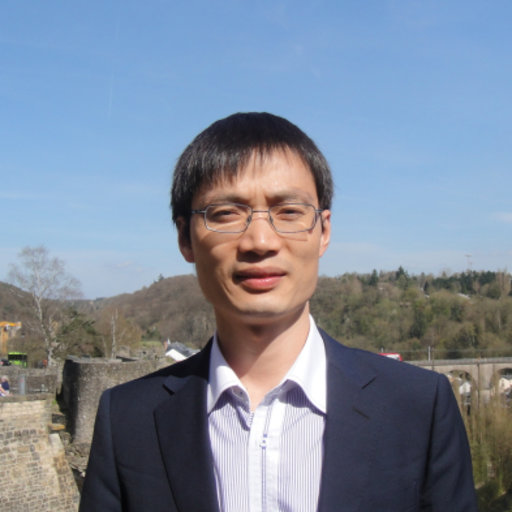}}]{Wenzhi Liao} (Senior Member, IEEE) received the B.Sc. degree in mathematics from Hainan Normal University, Haikou, China, in 2006, the Ph.D. degree in engineering from the South China University of Technology, Guangzhou, China, in 2012, and the Ph.D. degree in computer science engineering from Ghent University, Ghent, Belgium, in 2012.

From 2012 to 2019, he has worked as a Post- Doctoral Research Fellow first with Ghent University and then with the Research Foundation Flanders (FWO), Vlaanderen, Belgium. Since 2020, he has been with the Sustainable Materials Management, Flemish Institute for Technological Research (VITO), Mol, Belgium. He also works as the Guest Professor of Ghent University. His research interests include image processing, pattern recognition, and remote sensing. In particular, his interests include mathematical morphology, multitask feature learning, multisensor data fusion, and hyperspectral image (HSI) restoration.

Dr. Liao was a recipient of the Best Paper Challenge Awards in both the 2013 IEEE GRSS Data Fusion Contest and the 2014 IEEE GRSS Data Fusion Contest. He serves as an Associate Editor for the IEEE JOURNAL OF SELECTED TOPICS IN APPLIED EARTH OBSERVATIONS AND REMOTE SENSING (JSTARS).
\end{IEEEbiography}

\vskip -2\baselineskip plus -1fil
\begin{IEEEbiography}[{\includegraphics[width=1in,height=1.25in,clip,keepaspectratio]{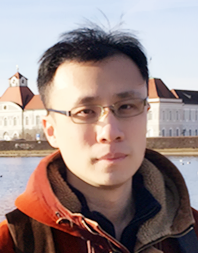}}]{Danfeng Hong}
(S'16-M'19) received the M.Sc. degree (summa cum laude) in computer vision, College of Information Engineering, Qingdao University, Qingdao, China, in 2015, the Dr. -Ing degree (summa cum laude) in Signal Processing in Earth Observation (SiPEO), Technical University of Munich (TUM), Munich, Germany, in 2019. 

Since 2015, he also worked as a Research Associate at the Remote Sensing Technology Institute (IMF), German Aerospace Center (DLR), Oberpfaffenhofen, Germany. Currently, he is research scientist and leads a Spectral Vision group at IMF, DLR, and also an adjunct scientist in GIPSA-lab, Grenoble INP, CNRS, Univ. Grenoble Alpes, Grenoble, France.

His research interests include signal / image processing and analysis, hyperspectral remote sensing, machine / deep learning, artificial intelligence and their applications in Earth Vision.
\end{IEEEbiography}

\vskip -2\baselineskip plus -1fil
\begin{IEEEbiography}[{\includegraphics[width=1in,height=1.25in,clip,keepaspectratio]{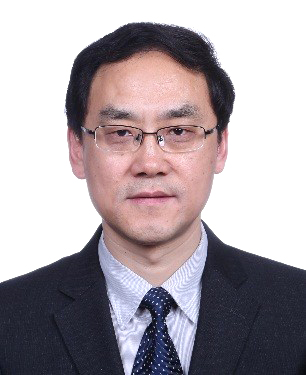}}]{Bing Zhang} (M'11–SM'12-F'19) received the B.S. degree in geography from Peking University, Beijing, China, in 1991, and the M.S. and Ph.D. degrees in remote sensing from the Institute of Remote Sensing Applications, Chinese Academy of Sciences (CAS), Beijing, China, in 1994 and 2003, respectively.

Currently, he is a Full Professor and the Deputy Director of the Aerospace Information Research Institute, CAS, where he has been leading lots of key scientific projects in the area of hyperspectral remote sensing for more than 25 years. His research interests include the development of Mathematical and Physical models and image processing software for the analysis of hyperspectral remote sensing data in many different areas. He has developed 5 software systems in the image processing and applications. His creative achievements were rewarded 10 important prizes from Chinese government, and special government allowances of the Chinese State Council. He was awarded the National Science Foundation for Distinguished Young Scholars of China in 2013, and was awarded the 2016 Outstanding Science and Technology Achievement Prize of the Chinese Academy of Sciences, the highest level of Awards for the CAS scholars.

Dr. Zhang has authored more than 300 publications, including more than 170 journal papers. He has edited 6 books/contributed book chapters on hyperspectral image processing and subsequent applications. He is the IEEE fellow and currently serving as the Associate Editor for IEEE Journal of Selected Topics in Applied Earth Observations and Remote Sensing. He has been serving as Technical Committee Member of IEEE Workshop on Hyperspectral Image and Signal Processing since 2011, and as the president of hyperspectral remote sensing committee of China National Committee of International Society for Digital Earth since 2012, and as the Standing Director of Chinese Society of Space Research (CSSR) since 2016. He is the Student Paper Competition Committee member in IGARSS from 2015-2019.
\end{IEEEbiography}

\vskip -2\baselineskip plus -1fil
\begin{IEEEbiography}[{\includegraphics[width=1in,height=1.25in,clip,keepaspectratio]{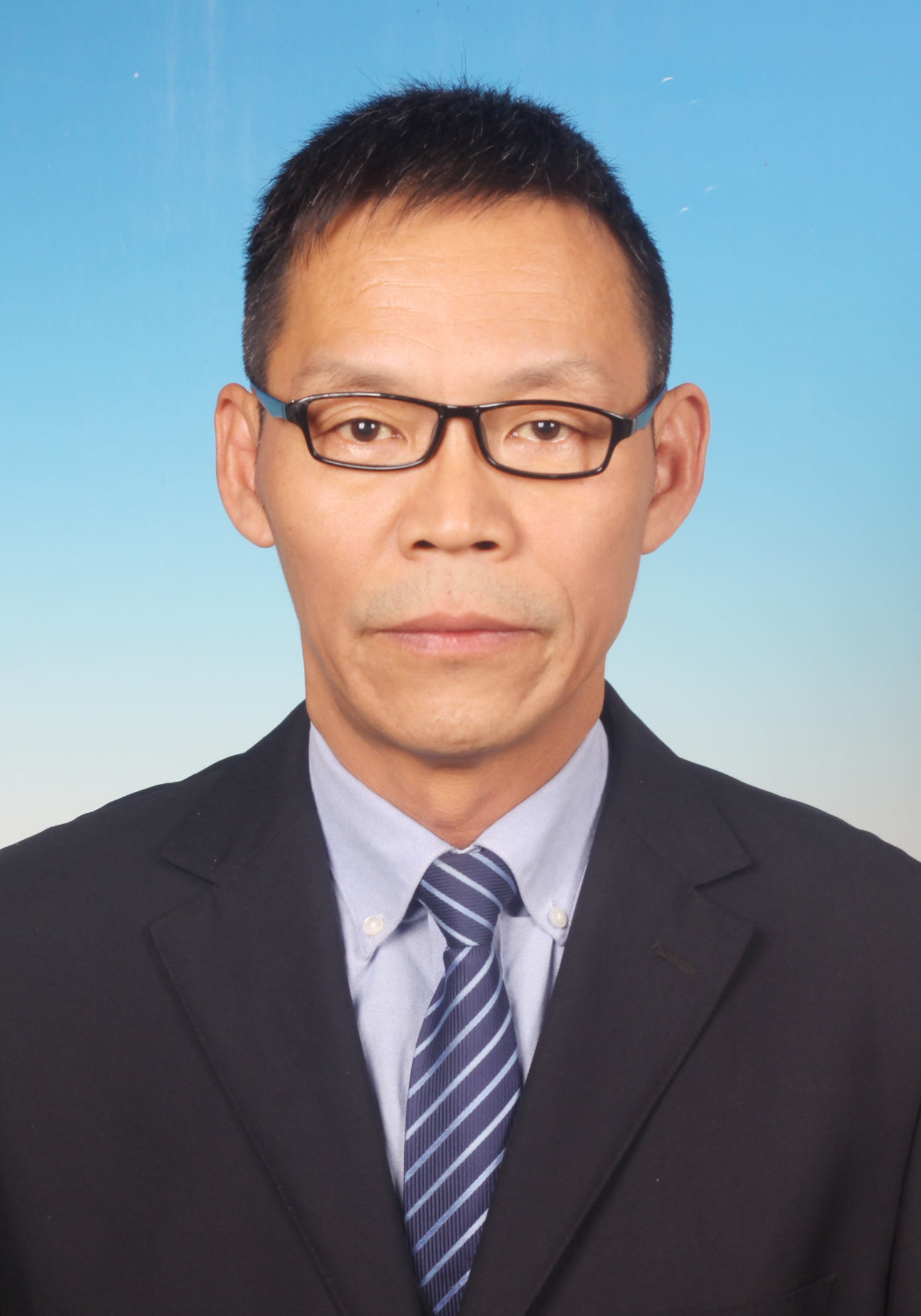}}]{Ximin Cui} received the B.S. degree in mine surveying from  China University of Mining and Technology, Xuzhou, China, in 1990, the M.S. degree in engineering surveying from China University of Mining and Technology, Beijing, China, in 1993, and the Ph.D. degree in mine engineering mechanics from China University of Mining and Technology, Beijing, China, in 1996. 

He is currently a Professor with the College of Geosciences and Surveying Engineering, China University of Mining and Technology (Beijing). He was the PI of 9 scientific research projects at national and ministerial levels, including projects by the National Natural Science Foundation of China (2000-2002, 2011-2013, 2015-2018) et al. He has published more than 170 peer-reviewed papers, and there are more than 70 papers included by Science Citation Index (SCI) and Engineering Village (EI). 

His research interests include mining subsidence, deformation monitoring and industrial surveying.  
\end{IEEEbiography}

\vskip -2\baselineskip plus -1fil
\begin{IEEEbiography}[{\includegraphics[width=1in,height=1.25in,clip,keepaspectratio]{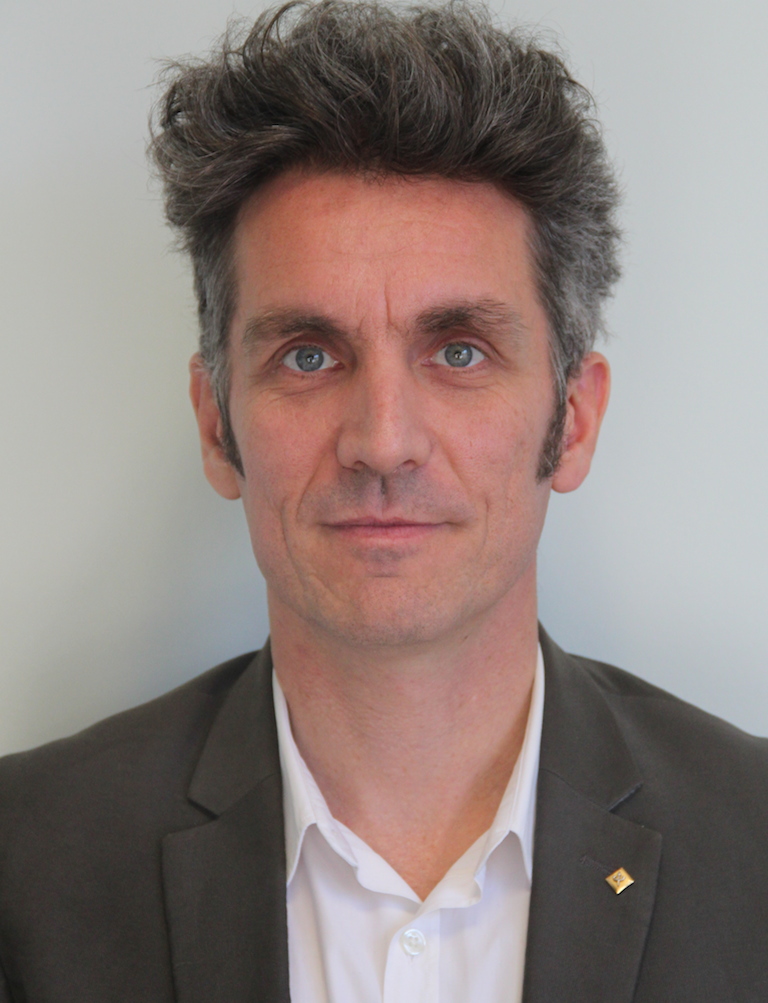}}]{Jocelyn Chanussot}
(M'04–SM'04-F'12) received the M.Sc. degree in electrical engineering from the Grenoble Institute of Technology (Grenoble INP), Grenoble, France, in 1995, and the Ph.D. degree from the Université de Savoie, Annecy, France, in 1998. Since 1999, he has been with Grenoble INP, where he is currently a Professor of signal and image processing. His research interests include image analysis, hyperspectral remote sensing, data fusion, machine learning and artificial intelligence. He has been a visiting scholar at Stanford University (USA), KTH (Sweden) and NUS (Singapore). Since 2013, he is an Adjunct Professor of the University of Iceland. In 2015-2017, he was a visiting professor at the University of California, Los Angeles (UCLA). He holds the AXA chair in remote sensing and is an Adjunct professor at the Chinese Academy of Sciences, Aerospace Information research Institute, Beijing.

Dr. Chanussot is the founding President of IEEE Geoscience and Remote Sensing French chapter (2007-2010) which received the 2010 IEEE GRS-S Chapter Excellence Award. He has received multiple outstanding paper awards. He was the Vice-President of the IEEE Geoscience and Remote Sensing Society, in charge of meetings and symposia (2017-2019). He was the General Chair of the first IEEE GRSS Workshop on Hyperspectral Image and Signal Processing, Evolution in Remote sensing (WHISPERS). He was the Chair (2009-2011) and  Cochair of the GRS Data Fusion Technical Committee (2005-2008). He was a member of the Machine Learning for Signal Processing Technical Committee of the IEEE Signal Processing Society (2006-2008) and the Program Chair of the IEEE International Workshop on Machine Learning for Signal Processing (2009). He is an Associate Editor for the IEEE Transactions on Geoscience and Remote Sensing, the IEEE Transactions on Image Processing and the Proceedings of the IEEE. He was the Editor-in-Chief of the IEEE Journal of Selected Topics in Applied Earth Observations and Remote Sensing (2011-2015). In 2014 he served as a Guest Editor for the IEEE Signal Processing Magazine. He is a Fellow of the IEEE, a member of the Institut Universitaire de France (2012-2017) and a Highly Cited Researcher (Clarivate Analytics/Thomson Reuters, 2018-2019).
\end{IEEEbiography}

\end{document}